\documentclass[12pt,preprint]{aastex}

\begin{document}

\title{Spectropolarimetric investigation of the propagation of magnetoacoustic waves and shock formation in sunspot atmospheres}

\author{Rebecca Centeno$^1$, Manuel Collados$^1$ and Javier Trujillo Bueno$^{1,2}$  }
\affil{$^1$Instituto de Astrof\'\i sica de Canarias, 38205 La Laguna, Tenerife, Spain}
\affil{$^2$Consejo Superior de Investigaciones Cient\'\i ficas (Spain)}
\email{rce@iac.es, mcv@iac.es, jtb@iac.es}

\begin{abstract}

Velocity oscillations in sunspot umbrae have been measured simultaneously
in two spectral lines: the photospheric
Silicon {\sc i}  \hbox{10827 \AA} line and the chromospheric Helium 
{\sc i} \hbox{10830 \AA} multiplet. From the full Stokes inversion of
temporal series of spectropolarimetric observations we retrieved, 
among other parameters, the line of sight velocity
temporal variations at photospheric and chromospheric heights.
Chromospheric velocity oscillations show a three minute period with a clear 
sawtooth shape typical of propagating shock wave fronts. Photospheric velocity 
oscillations have basically a five minute period, although the power spectrum 
also shows a secondary peak in the three minute band which has proven to be 
predecessor for its chromospheric counterpart.
The derived phase spectra yield a value of the atmospheric cut-off frequency
around $4$ mHz and give evidence for the upward propagation of higher frequency
oscillation modes. The phase spectrum has been reproduced with a simple model 
of linear vertical propagation of slow magneto-acoustic waves in a stratified 
magnetized atmosphere 
that accounts for radiative losses through Newton's cooling law. The model
explains the main features in the phase spectrum, and allows us to compute 
the theoretical time delay between the photospheric and
chromospheric signals, which happens to have a strong dependence on frequency.
We find a very good agreement between this and the time delay obtained 
directly from the cross-correlation of
photospheric and chromospheric velocity maps filtered around the 6 mHz band.
This allows us to infer that the 3-minute power observed at chromospheric
heights comes directly from the photosphere by means of linear wave
propagation, rather than from non-linear interaction of 5-minute (and/or
higher frequency) modes.

\end{abstract}


\section{Introduction}

\noindent The study of the generation and propagation of waves
in the solar atmosphere is a hot topic of research in astrophysics, since it 
provides information about the atmospheric structure and dynamics 
(e.g., Lites 1992; Bogdan 2000; Socas-Navarro, Trujillo Bueno and Ruiz Cobo, 
2000; Bogdan and Judge 2006), while at the same time it helps 
us identify the key mechanisms of chromospheric and coronal 
heating.  In fact, acoustic and magnetic waves and magnetic field 
reconnection have been mentioned in the literature as the most promising
heating mechanism candidates (Alfv\'en 1947; Biermann 1948; 
Schwarzschild 1948; Parker 1979; Ulmschneider and Musielak 2003).

Historically, sunspot oscillations have been classified into 
three different groups (e.g. Lites 1992): 
(1) photospheric umbral oscillations, which have basically a
5-minute period with an average rms amplitude of 75 m$\rm{s^{-1}}$. These 
oscillations are the apparent response of the umbral photosphere to the
5-minute p-mode oscillations. (2) Chromospheric umbral oscillations with 
periods around 180 s and amplitudes of a few kilometers per second,
and (3) running penumbral waves, seen in $\rm{H_{\alpha}}$ as disturbances
propagating radially outwards from the umbra.
They all seem to be different manifestations of the same
dynamical global phenomenon, though (e.g., Rouppe van der Voort et al. 2003). 

\noindent Simultaneous time-series observations of various spectral lines that 
sample different regions of the solar atmosphere is one of the most useful 
techniques 
for studying wave propagation (e.g., the review by Lites 1992 and references 
therein). For instance, Lites (1986) could provide hints of shock wave 
formation via the Doppler shifts observed in the Stokes $I$ profiles of the 
He {\sc i} 10830 \AA\ multiplet. By measuring the phase 
difference of the oscillations in different spectral lines,
this author could also infer the upward propagation of waves in the frequency 
band around 6.5 mHz (Lites, 1984). Other pioneering investigations on this
topic are those by Kneer et al. (1981). 

\noindent In the last 35 years, since the first report on chromospheric umbral 
oscillations was made (Beckers and Tallant, 1969), many works have been
published on this subject, accompanied by nearly an equal number of
differing findings, conclusions and contradictions 
yielded by the literature in this time. We refer 
the reader to recent reviews (such as those by Bogdan 2000 and Bogdan \& Judge
2006) for a 
comprehensive overview of present knowledge of oscillatory phenomena in
sunspots, both from the theoretical and the observational points of view.

Nowadays, theoretical investigations on this topic are mainly 
carried out by means of detailed numerical simulations. For 
instance, the hydrodynamical
simulations of Carlsson \& Stein (1995) suggest that
acoustic shock waves in the internetwork regions of the
solar atmosphere intermittantly heat the plasma there,
but with an acoustic shock heating that is insufficient to explain
quantitatively the emission line cores observed in far-UV lines.
Similar numerical simulations have recently begun to be extended to strongly 
magnetized regions of the solar atmosphere, taking into account the coupling 
among various MHD wave modes (e.g., Stein et al. 2004), but much work remains 
to be done prior to reaching a level of realism for which it becomes 
reasonable to start contrasting computed Stokes profiles with 
spectropolarimetric observations. In this respect, one of the aims of 
this paper is to provide high-quality observational information on the 
phenomenon of oscillations in sunspot umbrae, based on full Stokes-vector 
IR spectropolarimetry in photospheric and chromospheric lines.

This paper is organized as follows: Observations, data redution and inversion 
techniques are presented in sections 2 and 3. For the analysis, we follow a 
similar approach to that of Lites (1984, 1986) but
measuring instead the full Stokes-vector of the photospheric
Silicon {\sc i}  \hbox{10827 \AA} line and of the chromospheric Helium 
{\sc i} \hbox{10830 \AA} multiplet. The analysis of the photospheric and 
chromospheric LOS velocities, and the relation between them, are shown in 
section 4. As we shall see below, we 
are able to provide very clear observational evidence for the upward 
propagation of waves from the photosphere to the chromosphere within the 
umbra of a sunspot, including an unprecedent measurement of the time
delay between the signals and the detection and characterization of the 
photospheric driving piston. A brief discussion can be found
in section 5, followed by some final remarks in section 6.

\section{Observations}

The observations analyzed in this paper were carried out at the German Vacuum 
Tower Telescope (VTT) of the Observatorio del Teide on October 1st 2000 and
May 9th 2001, 
using the Tenerife Infrared Polarimeter (TIP, Mart\'\i nez Pillet et al. 1999).
This instrument allows us to take simultaneous images of the four Stokes
parameters as a function of wavelength and position along the spectrograph 
slit, with a temporal sampling up to 0.5 seconds. In order to obtain a better
signal-to-noise ratio, several images were added up on-line, with a final 
temporal sampling of several seconds.  
The slit (0''.5 wide and 40'' long) was placed over the targets and
was kept fixed during the entire 
observing run (approx. 1 hour). The image
stability was obtained using a correlation tracker device (Ballesteros et al, 
1996) which compensates for the image motion induced by the Earth's high 
frequency atmospheric 
variability, as well as for solar rotation.

The observed spectral range spanned from 10825.5 to \hbox{10833 \AA}, with a 
spectral sampling of 31 m\AA\ per pixel. This spectral
region is a powerful diagnostic window of the solar atmospheric properties
since it contains valuable information coming from two different regions in the
atmosphere (Harvey \& Hall 1971; R\"uedi, Solanki, Livingston 1995; Trujillo 
Bueno et al. 2002; Trujillo Bueno et al. 2005; Solanki et al. 2003). 
It includes three spectral features. The first is 
a photospheric Si {\sc i} line at \hbox{10827.09 \AA}. Next to it lies
the chromospheric Helium {\sc i} \hbox{10830 \AA} line, which is indeed a triplet, 
whose blue component ($\lambda$ \hbox{10829.09 \AA}) is quite
weak and difficult to see in an intensity spectrum, and whose red components
($\lambda$ 10830.25, $\lambda$ \hbox{10830.34 \AA}) appear blended. 
The formation mechanism of this triplet is still not
fully understood, though it is thought to be generated in a thin layer in the
high chromosphere, about 2000 km above the base of the photosphere 
(Avrett et al. 1994).  
The third spectral feature is a water vapour line (R\"uedi et al. 1995) of 
telluric origin that can only be used for calibration purposes, since it 
generates no polarization signal. 

We chose two different target sunspots for the analysis presented
in this paper. On both occasions the slit was placed over the center of
the sunspot. Table \ref{tbl-1} shows the details for both data sets.

\section{Data reduction and inversion}

Flatfield and dark current measurements were performed at the beginning and
the end of both observing runs and, in order to compensate for the telescope
instrumental polarization, we also took a series of polarimetric calibration
images. The calibration optics (Collados 1999) allows us to obtain the 
Mueller matrix of the light path between the instrumental calibration
sub-system and the detector. This process leaves a section of the telescope
without being calibrated, so further corrections of the residual cross-talk
among Stokes parameters were done: the I to Q, U and V cross-talk was 
removed by forcing to zero the continuum polarization, and the circular
and linear polarization mutual
cross-talk was calculated by means of statistical techniques (Collados 2003). 

In order to infer the physical parameters of the magnetized atmosphere in 
which the measured
spectral lines were generated, we carried out the full Stokes inversion of both
the Silicon line and the Helium triplet within the umbra of the sunspot for
the whole time series of observations and for both data sets. 
Thus, we were able to obtain the temporal variability of several physical 
quantities (line of sight velocity, magnetic field intensity and 
orientation...) at the photospheric and chromospheric regions where the 
observed spectral line radiation originates, though in this paper we will 
concentrate only on the results concerning the line of sight velocity temporal 
fluctuations. 
We could have used a simpler method to infer Doppler velocities
rather than inversion techniques, but we decided to stick to the latter  
after comparing the results with those obtained from a preliminary analysis 
(in which we calculated velocities by measuring directly the position of the 
intensity minimum and the Stokes V zero-crossing), and finding that they
yielded very similar results.

\noindent The information encoded in the Silicon line radiation was retrieved 
by using the code LILIA, developed by 
Socas-Navarro (2001). LILIA is a package for the synthesis and inversion of
Stokes profiles induced by the Zeeman effect. It is based on the 
assumption of local thermodynamic equilibrium (LTE), and takes into account 
the Zeeman induced polarization pattern of the spectral lines. 
A guess atmosphere is iteratively modified by the code, using a 
Levenberg-Marquardt minimization algorithm (Press et al. 1988), until
the synthetic profiles mimic the observed ones, in a least square sense.
LILIA returns not only the values for the thermodynamic and physical 
parameters, but also their stratification in the atmosphere.
All the values for photospheric parameters presented from now on in 
this paper, refer to the height corresponding to $\log(\tau_{500}) = -2$.

\noindent The inversion of the Helium lines was carried out using a 
code based on the Milne-Eddington (ME) approximation, similar to that 
described by Socas-Navarro, Trujillo Bueno \& Landi Degl'Innocenti (2004). 
We decided to give no weight to the blue
component of the triplet in the inversion because it seemed to be
contaminated with some other unknown spectral feature, whose possible
physical origin is discussed in Centeno et al. (2005). 
Although the ME approach does not provide the 
stratification of the atmospheric parameters, it returns precise velocity 
values and
an average (over the region of formation) for the rest of the magnitudes
(see, e.g., Westendorp Plaza et al. 1997).

\section{Analysis}

\subsection{Shock waves}

\begin{figure}
\begin{center}
\includegraphics[angle=0,scale=0.5]{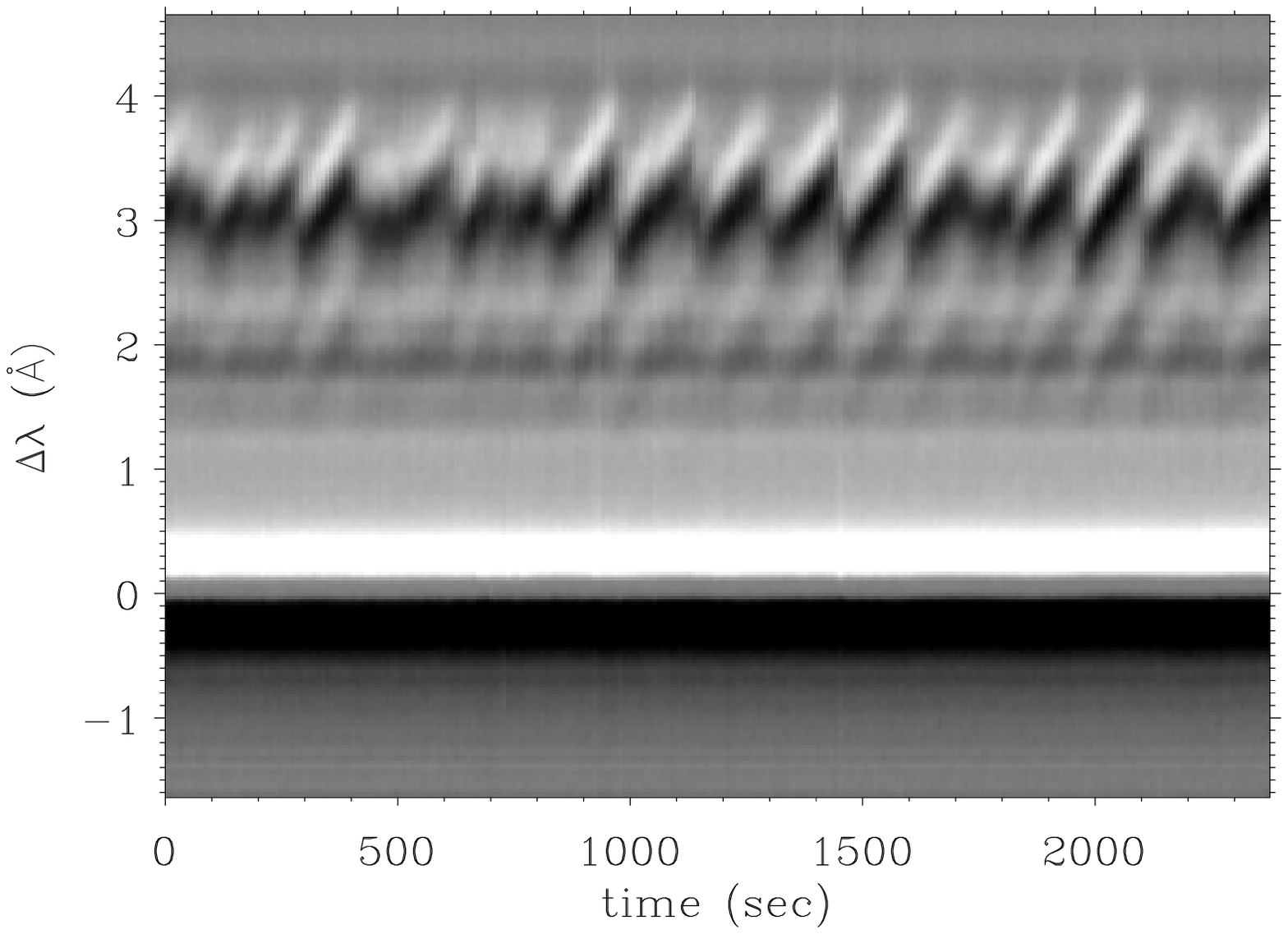}
\caption{Temporal evolution of Stokes V. The horizontal axis represents time, increasing to the right, and the vertical axis wavelength (with origin at the position of the Silicon rest wavelength). The Silicon Stokes V profile (lower part of the figure) shows no apparent change with time in this presentation, while Helium profiles (upper part) show periodic Doppler shifts with a clear saw-tooth shape.}
\label{fig:stokes_v_map}
\end{center}
\end{figure}

Figure \ref{fig:stokes_v_map} shows the temporal evolution of the Stokes V
profiles for one position inside the umbra of sunspot \#1 obtained
directly from the observations. The horizontal and vertical axes represent 
time and wavelength, respectively (the origin of wavelength scale being the 
rest position of the Silicon line).
Around the zero-wavelength position we see the Silicon
Stokes V profile, which at first sight does not seem to change with time. 
On the other hand, the upper part of the figure shows the evolution of the 
circular polarization observed in the Helium multiplet, with the
high contrasted red components on top (at $\sim$ 3.3 \AA) and the blue 
component rather weak, underneath (at $\sim$ 2 \AA). 
Both signals exhibit the same behaviour: the positions of the 
zero-crossing of the Stokes profiles show periodic shifts in wavelength
with a clear sawtooth shape (i.e. sudden blue-shifts followed by 
slower red-shifts), suggestive of shock wave trains.

\begin{figure}
\begin{center}
\includegraphics[angle=0,scale=0.4]{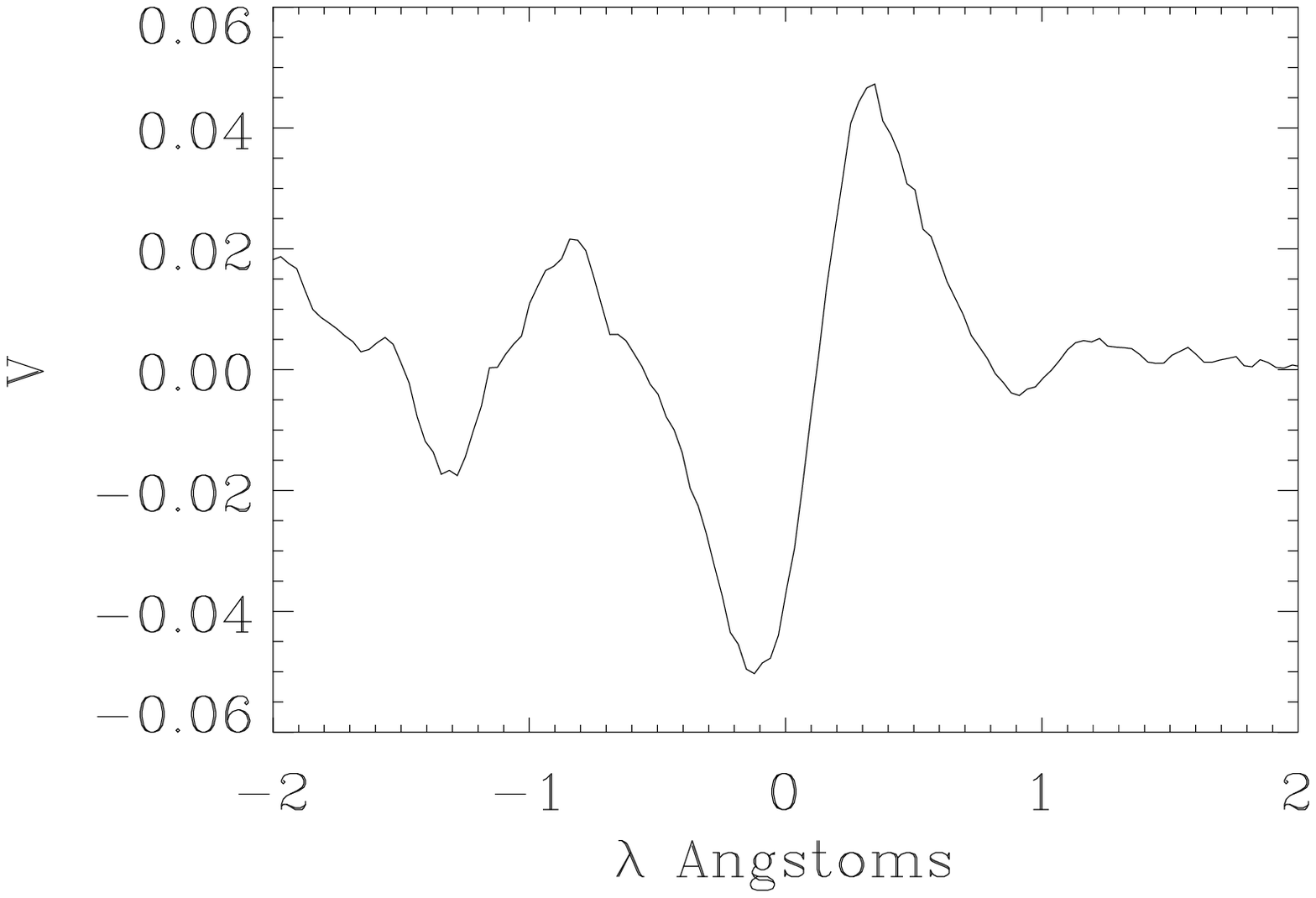}\\
\includegraphics[angle=0,scale=0.4]{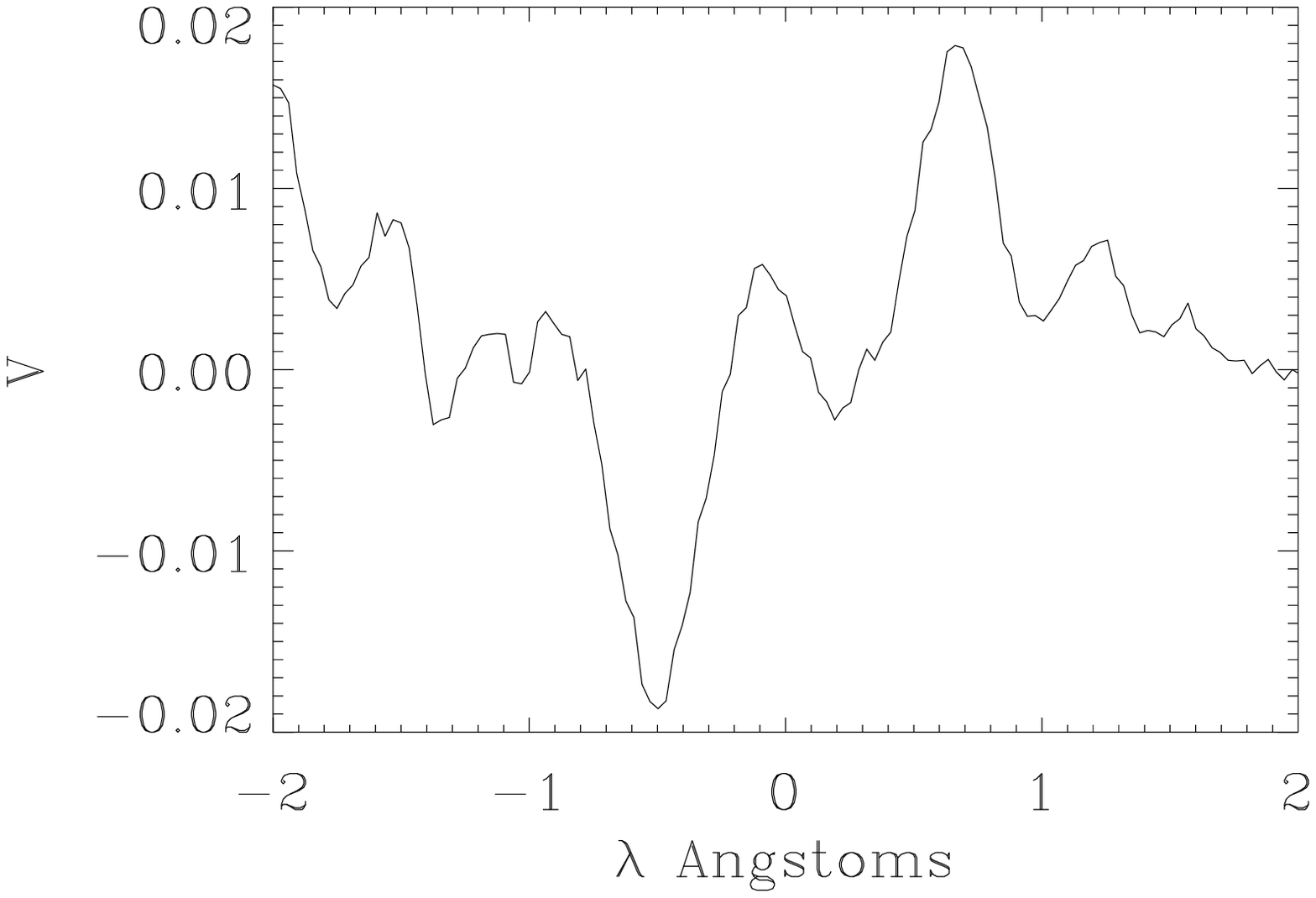}
\caption{{\bf Top:} Normal Stokes V profile while the line is undergoing
a redshift. {\bf Bottom:} Complex shaped Stokes V profile while the line
is undergoing a blueshift.}
\label{fig:stokesvprof}
\end{center}
\end{figure}

\noindent It is interesting to note that the Stokes V profiles show irregular 
shapes during blue-shifts, suggesting they are not resolved within the 
temporal and/or spatial sampling of our observations or even that they are
simply the result of integration along the line-of-sight in a shocked plasma. 
This happens not
only for sunspot \#1 (with a 7.9s sampling) but also for sunspot \#2
(with a much finer sampling of 2.1s).
Figure \ref{fig:stokesvprof} shows two different Stokes V profiles 
for the Helium triplet, corresponding to a unique position inside the umbra
of sunspot \#1. The two profiles were measured at different times: while
the one on top corresponds to a snapshot of a redshift, 
the one on the bottom was measured while the Helium line was undergoing a 
blueshift.

\subsection{Results from the inversions}

After carrying out the full Stokes inversion of the Silicon line and
the Helium triplet inside the umbrae of sunspots \#1 and \#2, 
we obtained the temporal variation of the atmospheric structure at the
photosphere and the chromosphere, for all the positions along the slit.
We will focus on the results for the line of sight velocity in this paper.
As mentioned earlier, the analysis of the remaining retrieved parameters 
will be left for a subsequent paper.

\begin{figure}
\begin{center}
\includegraphics[angle=0,scale=0.5]{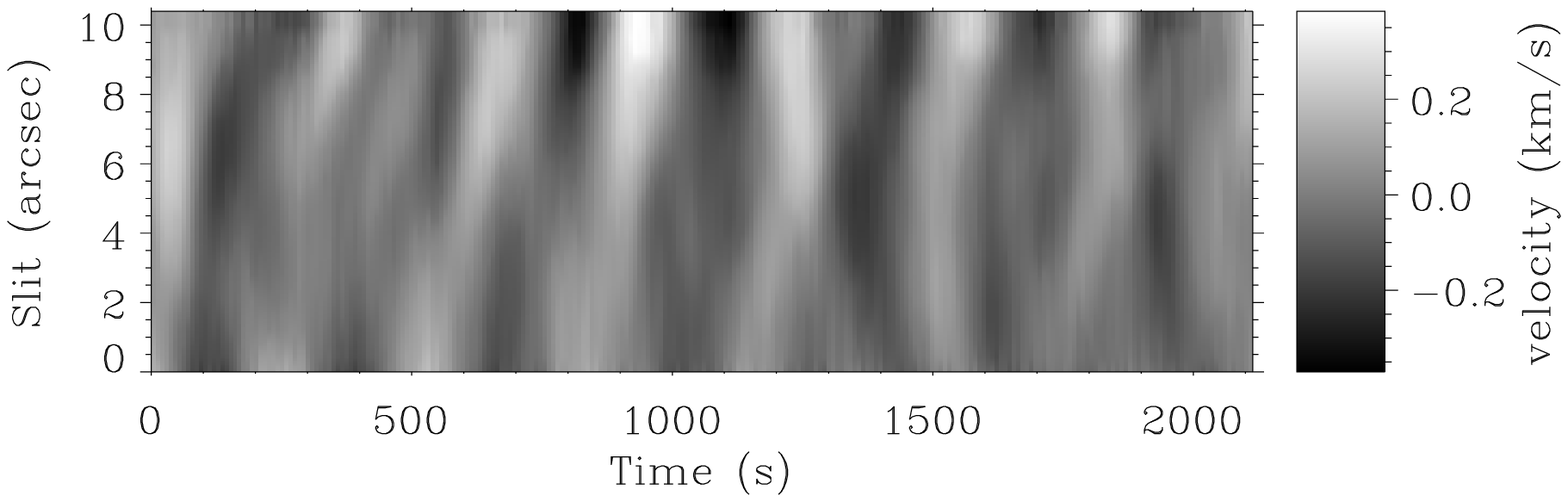}
\includegraphics[angle=0,scale=0.5]{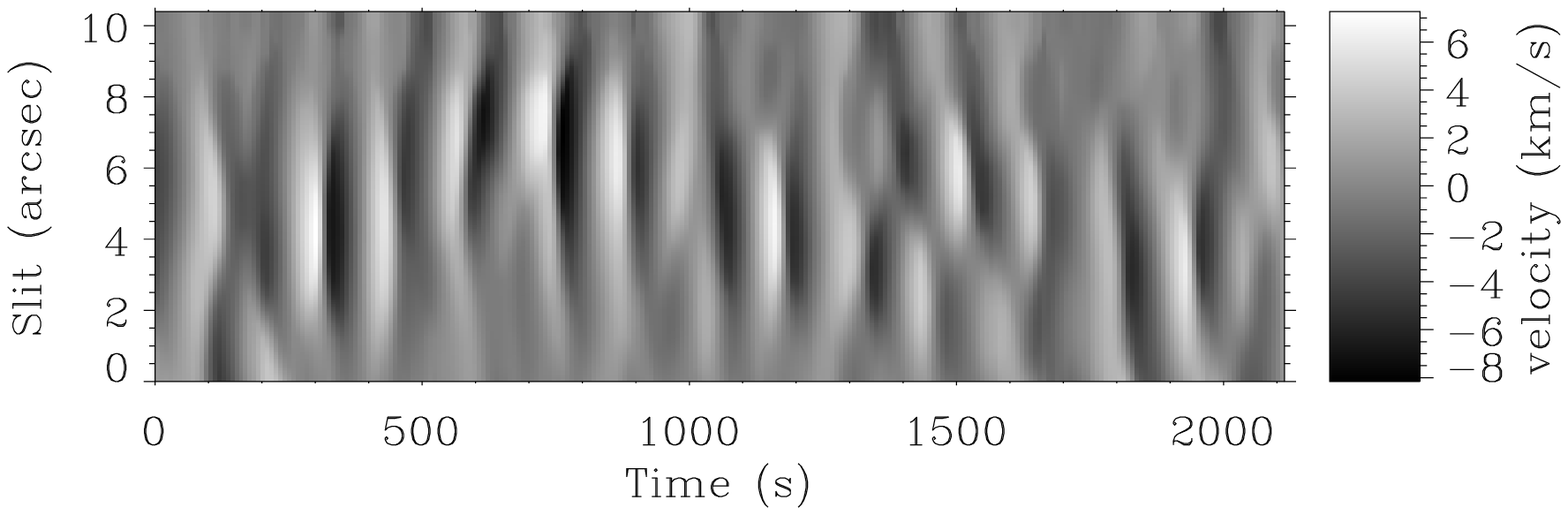}
\caption{Velocity maps for the photosphere (above) and the chromosphere
(below) inside the umbra of sunspot \# 2. Horizontal axis represents the
time (in seconds) and vertical axis the position along the slit (arcsecs).
Black means negative velocity (material approaching the observer).}
\label{fig:velocitymaps}
\end{center}
\end{figure}

Figure \ref{fig:velocitymaps} shows the photospheric (above) and 
chromospheric (below) line of sight velocity maps for the umbra of
sunspot \#2, obtained from the inversions of the Silicon and Helium lines 
respectively. The horizontal and vertical axes 
represent time and position along the slit. Black means
negative velocity (matter approaching the observer). In both maps, we have 
subtracted a slow linear variation of the velocity due to the Sun-Earth 
relative motion.
The phase of the oscillations changes slowly accross the umbra in 
the chromosphere, while in the photosphere this 
variation is even slower, and the oscillations seem to keep the coherence
over larger regions.
The typical size of a photospheric patch is 5 to 10 arcsec while the size 
of a chromospheric patch is between 2 and 5 arcsec.

\begin{figure}
\begin{center}
\includegraphics[angle=0,scale=0.3]{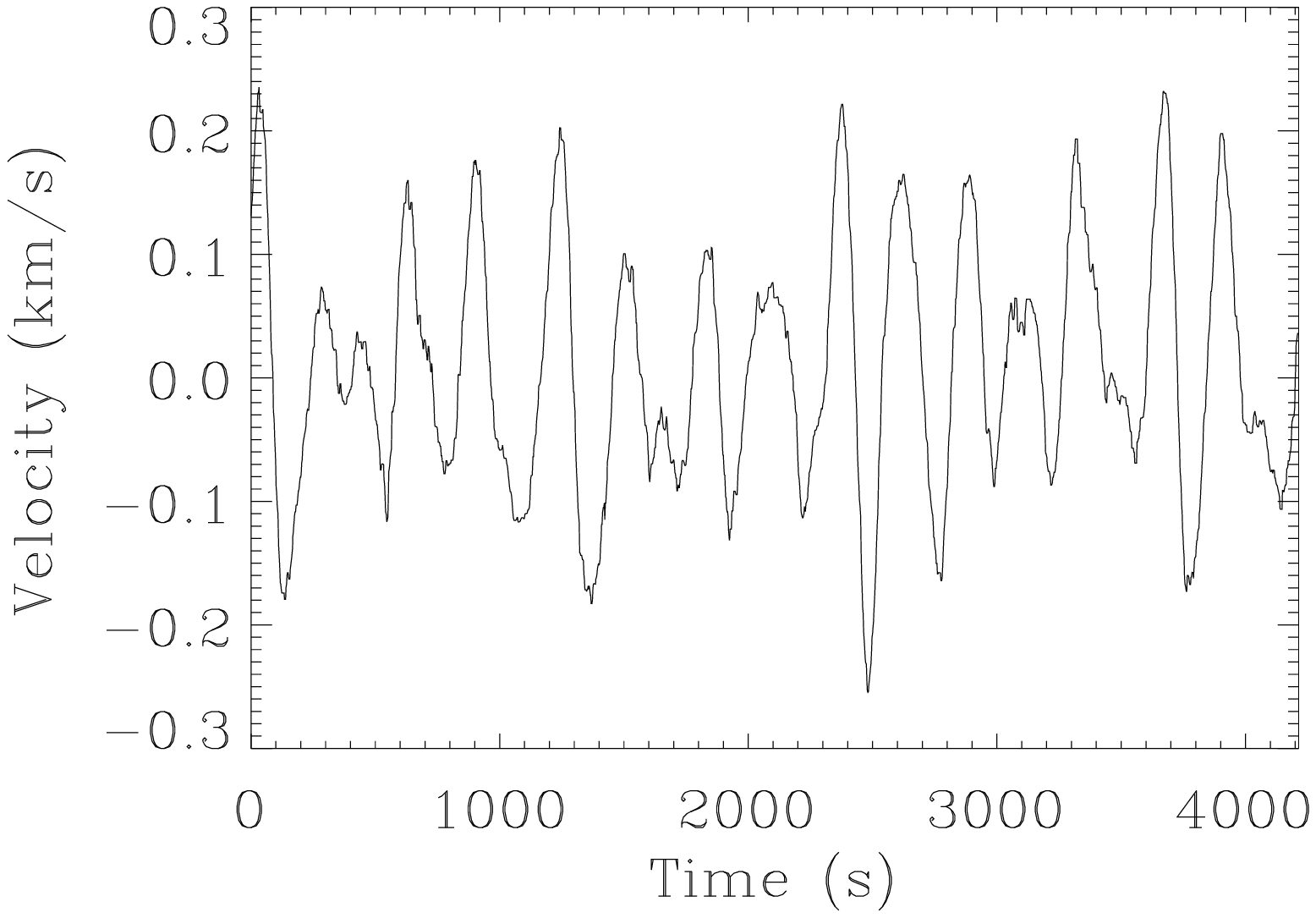}
\includegraphics[angle=0,scale=0.3]{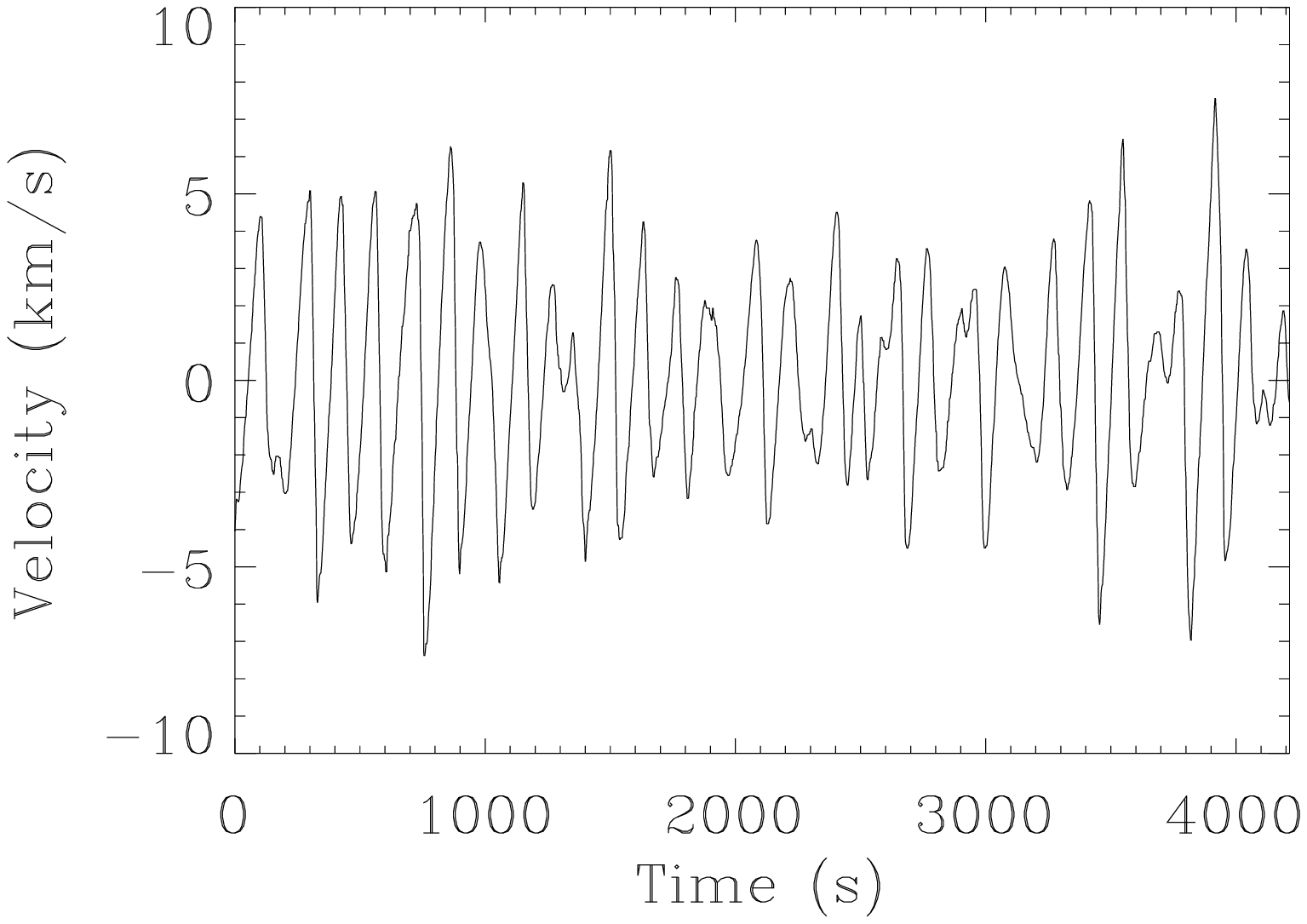}
\includegraphics[angle=0,scale=0.3]{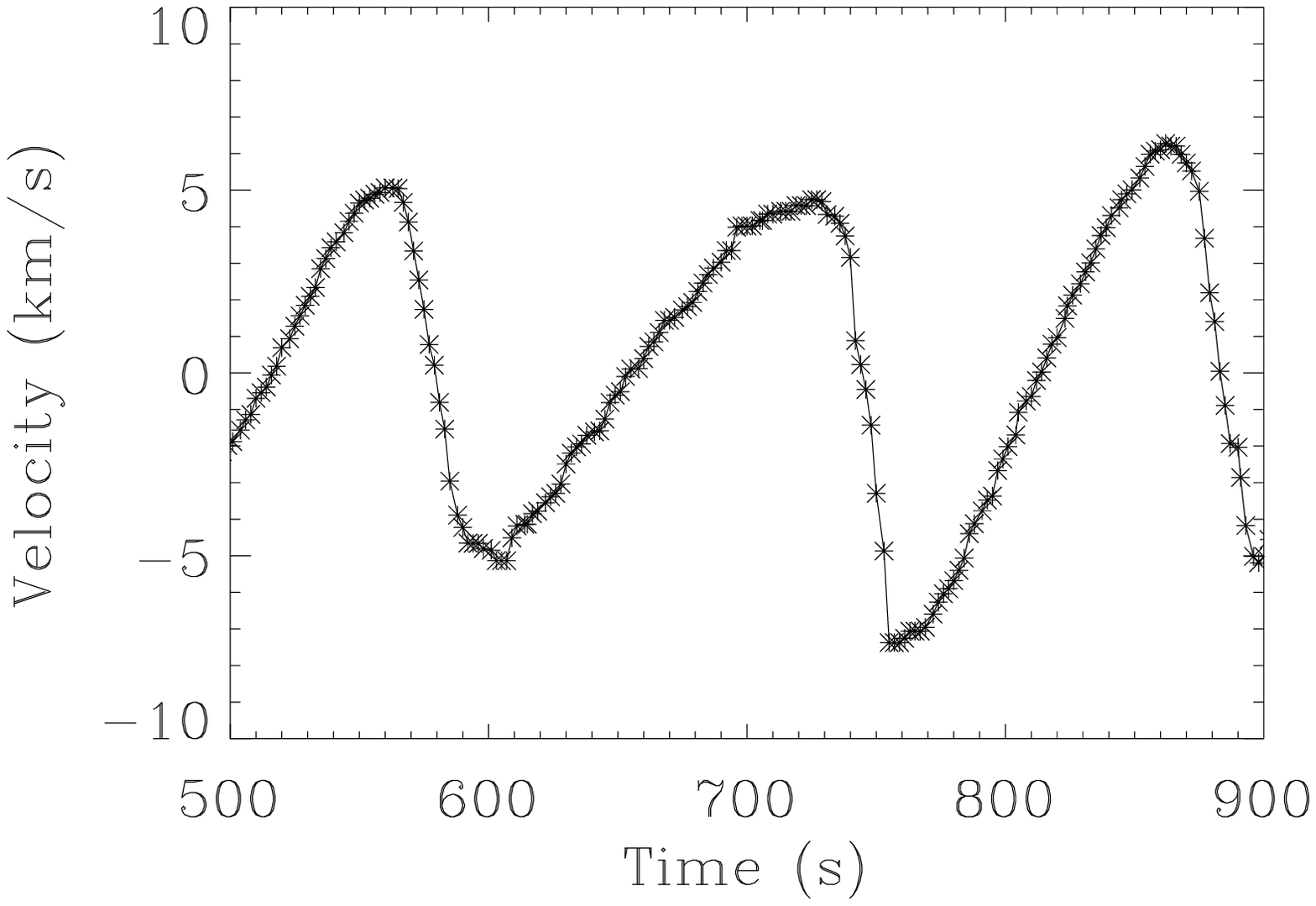}
\caption{Photospheric {\bf (left)} and chromospheric {\bf (middle)} line of 
sight velocity signals as a function of time, for one point of the slit 
inside the umbra of sunspot \#2.
{\bf Right}: Detail of chromospheric velocity at one point of the slit. The stars represent the measured values. The sawtooth shape is clearly seen, with a slow evolution of the velocity towards positive values followed by a sudden blue-shift, indicating the presence of shock waves
}
\label{fig:velocityslit}
\end{center}
\end{figure}

Figure \ref{fig:velocityslit} shows the temporal evolution of the 
photospheric (left) and chromospheric (middle) line of sight velocities
for one position inside the umbra of sunspot \#2.
The photospheric velocity, with an rms noise level of about 30 m$\rm{s^{-1}}$, 
shows an approximately sinusoidal pattern with a peak to peak amplitude of 
400 ms$^{-1}$ and a period within the five minute band. 
On the other hand, chromospheric velocity oscillations (shown in more
detail in the right panel of Figure \ref{fig:velocityslit}) have a well 
defined three minute period and quite large peak to peak variations
(of the order of 10-15 kms$^{-1}$). Also, the oscillation pattern has a clear 
sawtooth shape that indicates the presence of shock wave trains at 
chromospheric heights, as already mentioned in the former section.

\subsection{Power spectra.}

\begin{figure}[ht]
\begin{center}
\includegraphics[angle=0,scale=0.4]{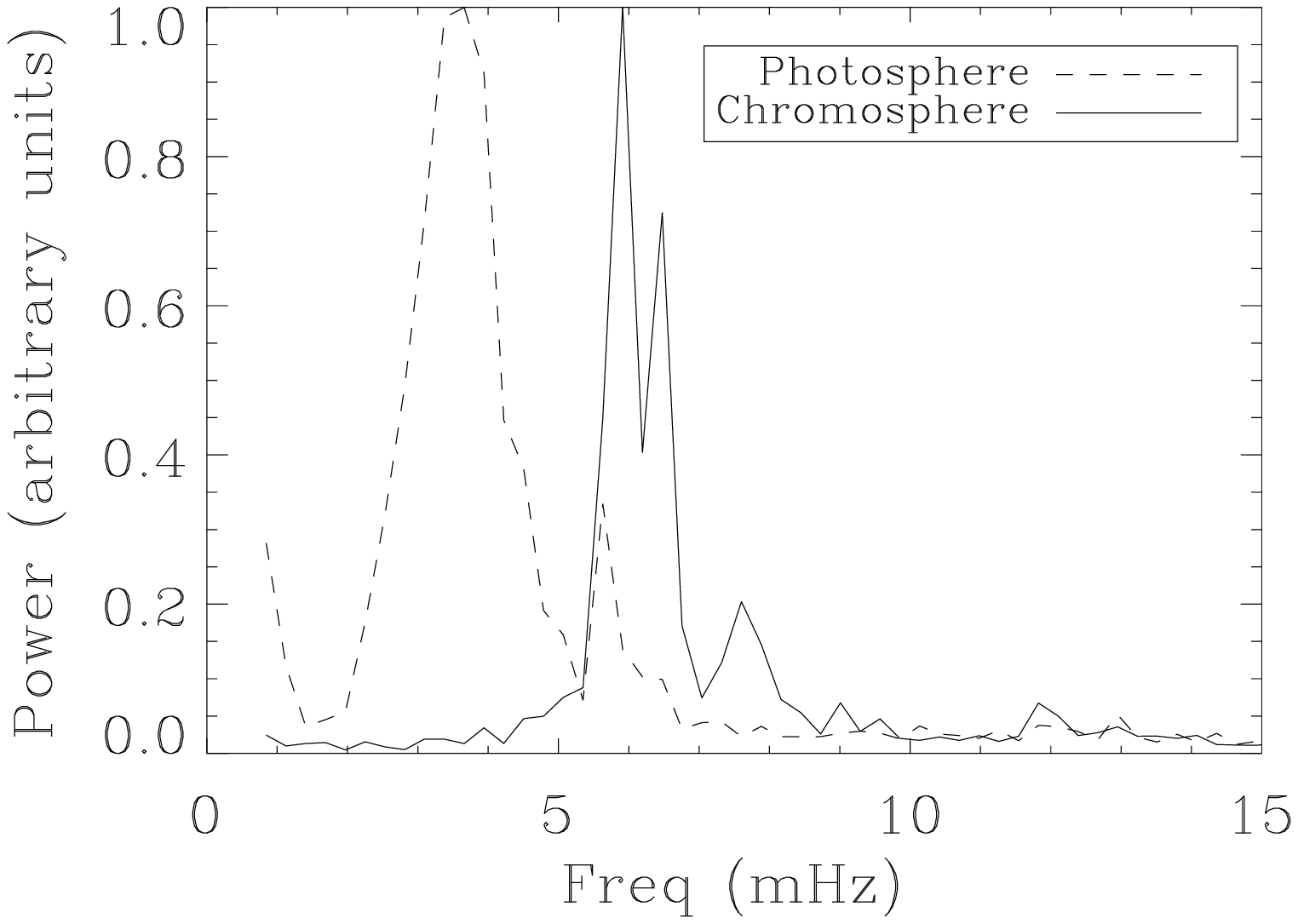}\\
\includegraphics[angle=0,scale=0.4]{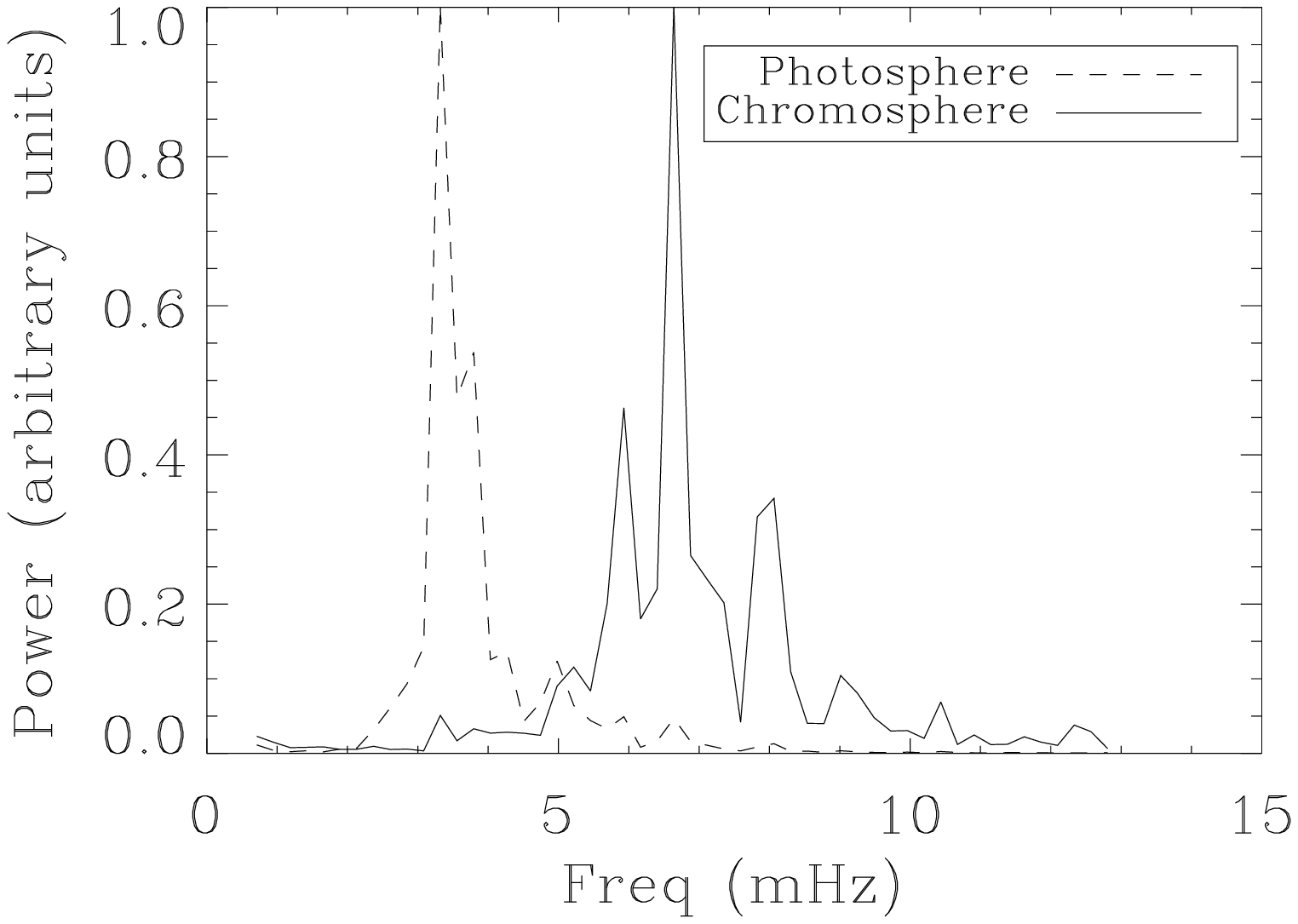}
\caption{Average umbral power spectra for sunspot \#1 (top) and for sunspot \#2 (bottom). {\bf Solid}: Power spectrum of the chromospheric velocity oscillations averaged over the umbra, with a peak around 6 mHz. {\bf Dashed}: Photospheric velocity power spectrum averaged over the entire umbra, with a peak around 3.3 mHz and secondary peaks around 6 mHz.}
\label{fig:power_spectra}
\end{center}
\end{figure}

\noindent Figure \ref{fig:power_spectra} shows the power spectra averaged 
over the entire umbra of sunspot \#1 (top) and sunspot \#2 (bottom), for 
both the 
chromospheric (solid) and the photospheric (dashed) velocity signals.
In the chromospheric signal, power is concentrated
between 5 and 8 mHz (3 minute band) with a clear peak around 6 mHz. Note 
that there is nearly no power at all in the 5 minute band ($\sim$ 3.3 mHz).
On the other hand, the photospheric velocity power spectrum has most of 
its power concentrated in the range from 2 to 5 mHz, which corresponds to 
the well known five minute oscillations. But there are also secondary peaks
between 5 - 7 mHz which we believe to be the photospheric counterpart of 
the chromospheric 3-minute oscillations, as will be shown further on.

\subsection{Phase diagrams}

\begin{figure}[ht]
\begin{center}
\includegraphics[angle=0,scale=0.35]{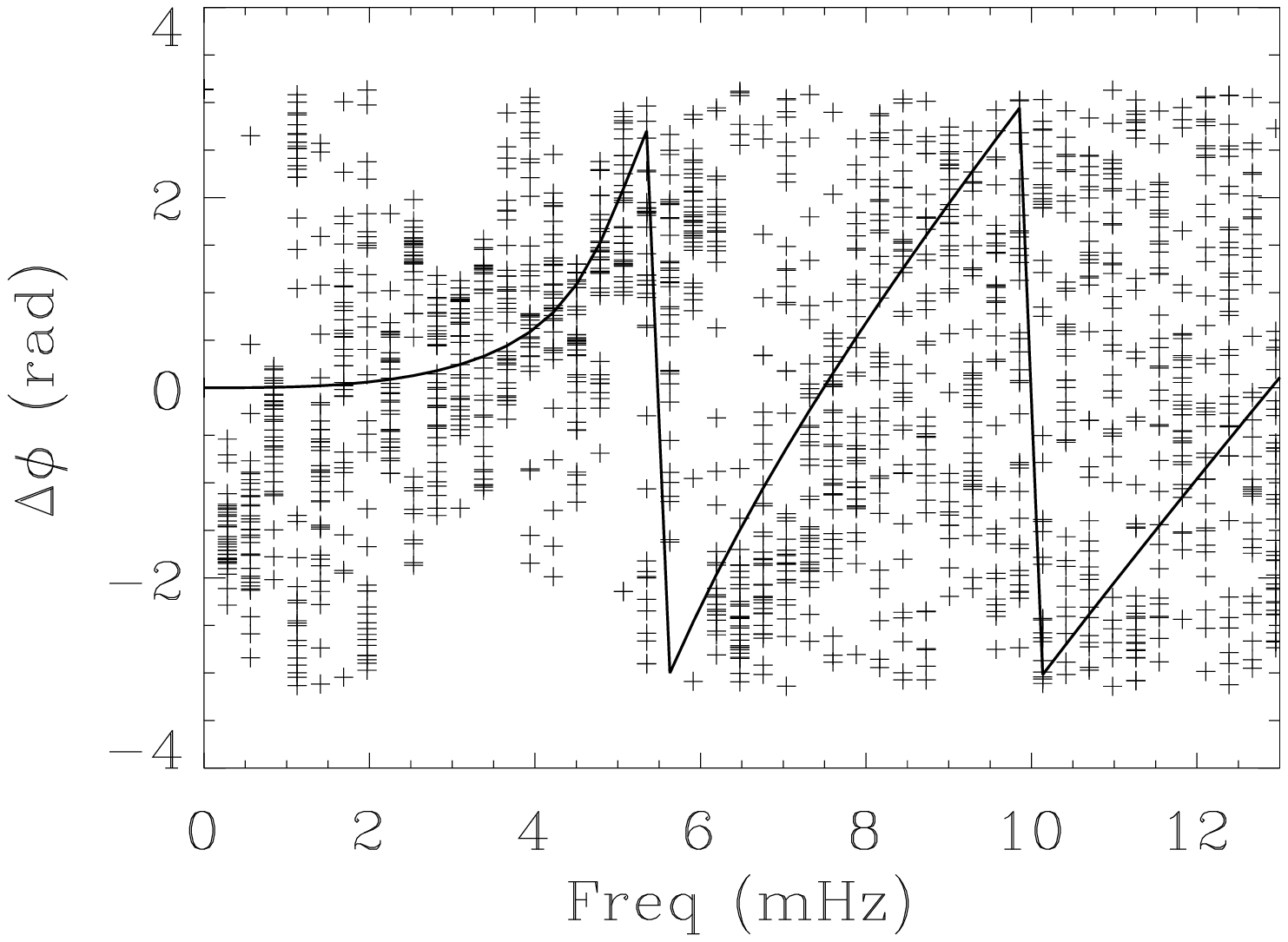}
\includegraphics[angle=0,scale=0.35]{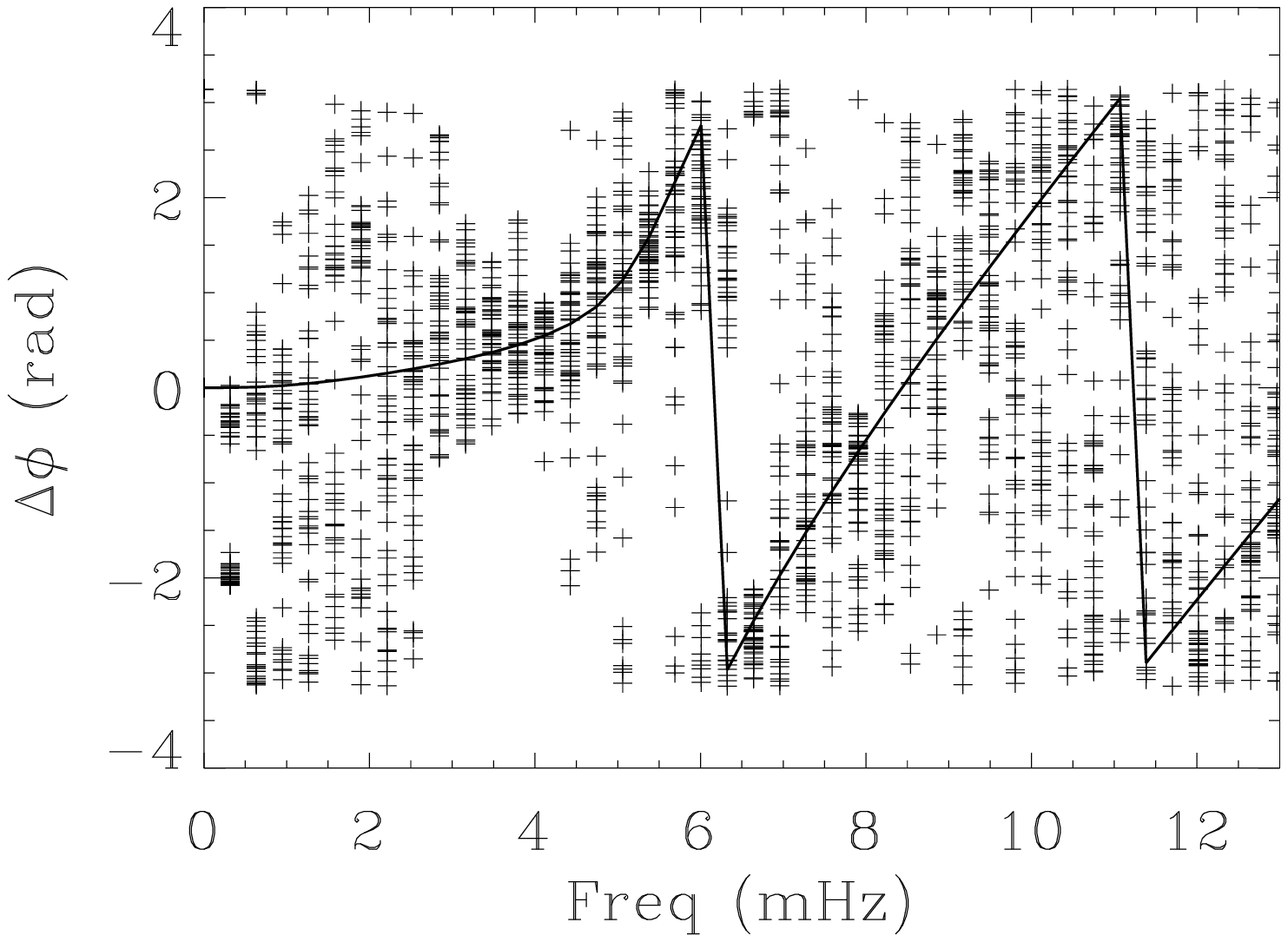}\\
\includegraphics[angle=0,scale=0.35]{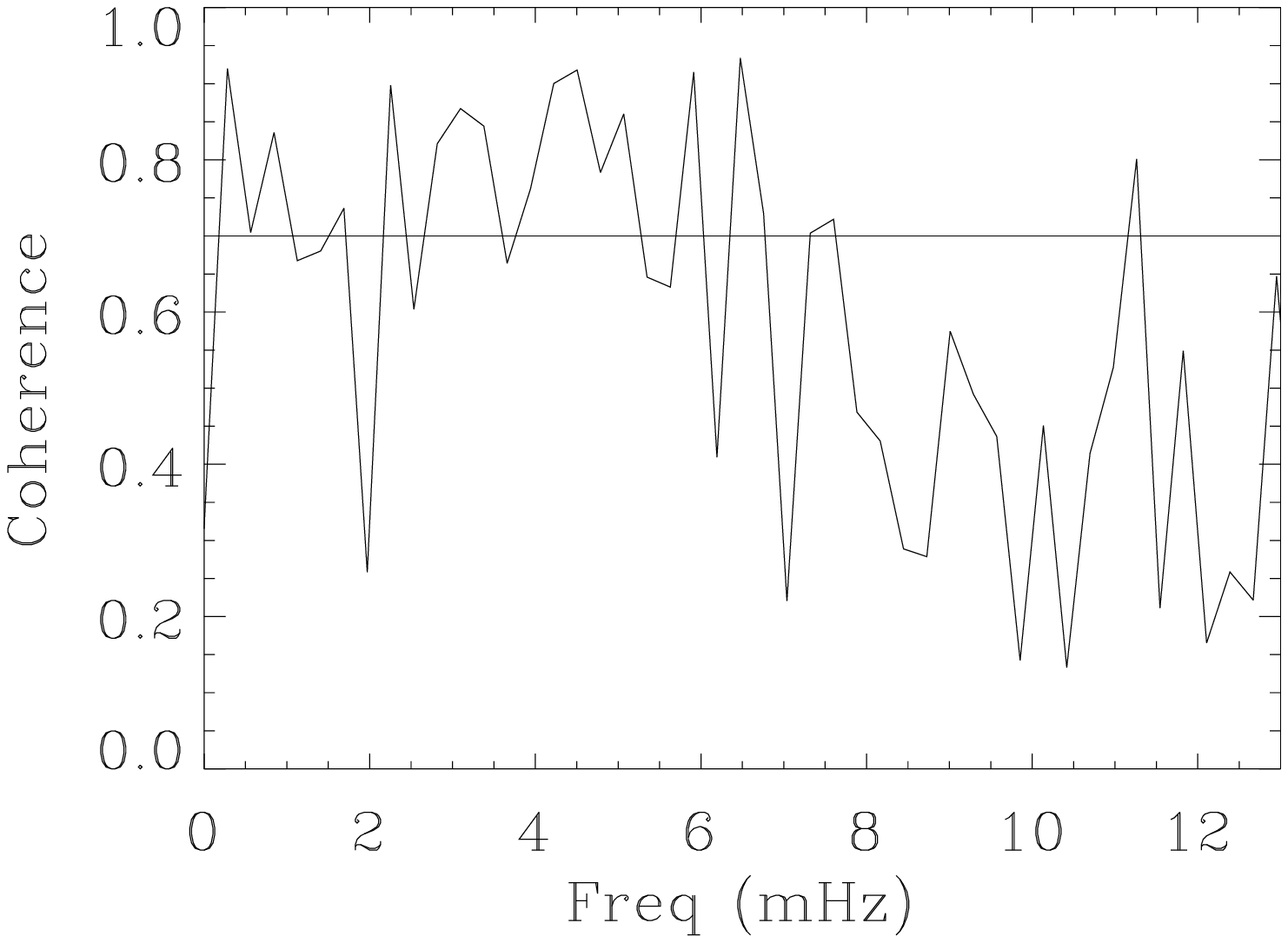}
\includegraphics[angle=0,scale=0.35]{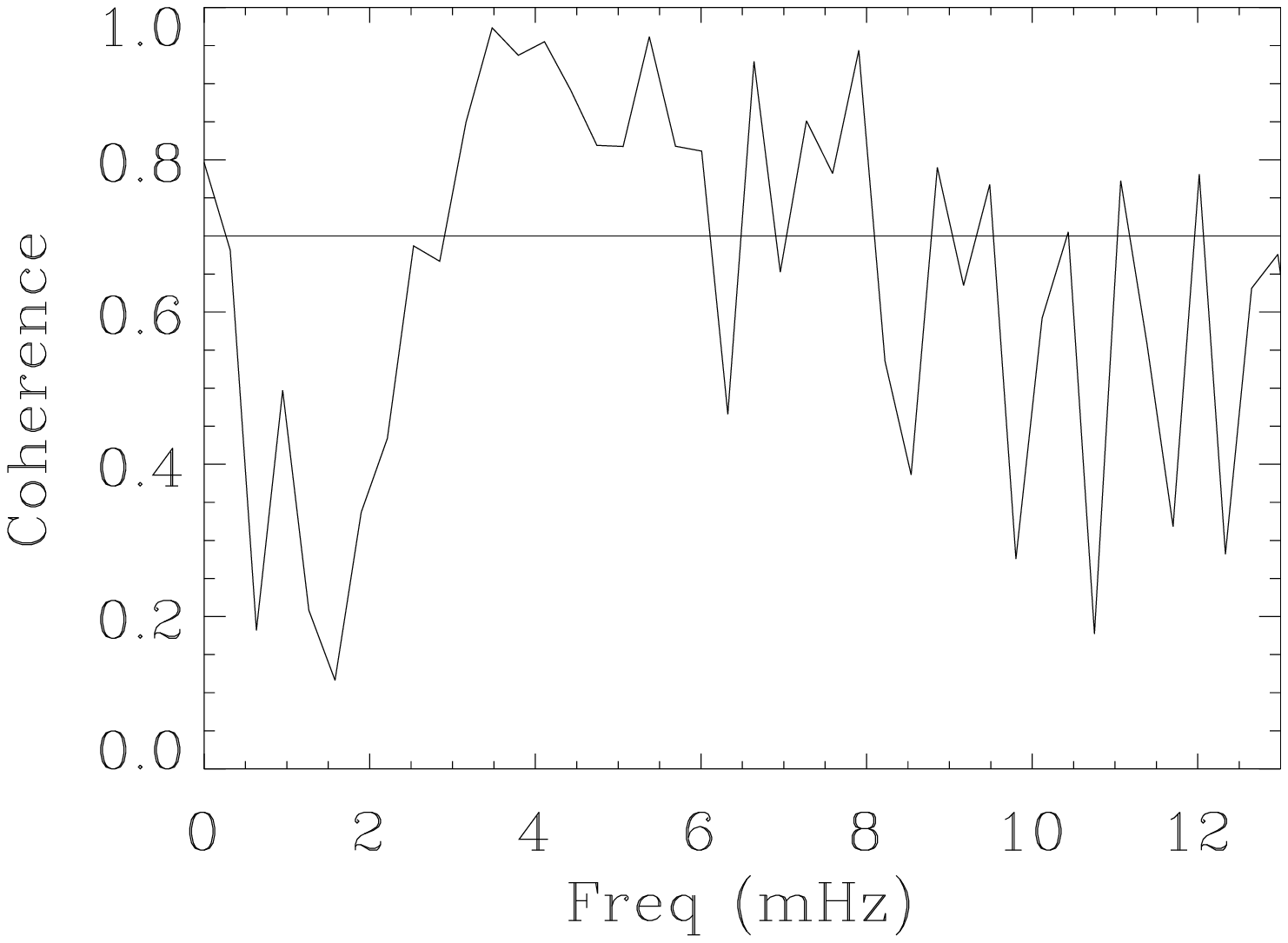}
\caption{{\bf Above:} Phase spectra for all the points in the umbra of 
sunspot \#1 (left) and sunspot \#2 (right). On the x-axis the frequency 
(mHz). On the y-axis, the phase difference between the Fourier transform of 
the chromospheric velocity oscillation and the photospheric velocity 
oscillation (radians). The solid line represents the best fit from the 
theoretical model. 
{\bf Below:} Coherence spectra for both cases. It gives 
information about the validity of the phase spectra. Horizontal line sets
the confidence limit (coherence greater than 0.7) above which, we can 
believe the information given by the phase spectrum.}
\label{fig:phase_spectra}
\end{center}
\end{figure}

The upper part of Figure \ref{fig:phase_spectra} shows the phase difference 
($\Delta\phi$) between the chromospheric and the photospheric 
velocity signals, as a function of frequency.
On the left, the results for umbra \#1, and on the right for umbra \#2.
Each cross on the figure is obtained as the difference from the 
chromospheric phase and the photospheric phase for one frequency and 
one position inside the umbra of the sunspot.
The phases are obtained directly from the Fourier transform of the velocity
signals, for each position along the slit. Note that there is a $2\pi$
indetermination in the computation of the phase value, so the phase
difference will be cyclic every $2\pi$.

\noindent The lower part of Figure \ref{fig:phase_spectra} shows the coherence
spectra for both data sets. Coherence spectrum is intimately related to
phase spectrum and it tells us whether the phase difference between two 
signals for one harmonic $\omega$ is characteristic of the signals
or, on the contrary, is an arbitrary feature. For this reason, coherence
is a statistical definition, and makes no sense when calculated
between two velocity signals measured at two different heights and the
same spatial position.
The horizontal line delimits the confidence limit at 0.7, above 
which we
consider the coherence is significant, and the information given by
the phase spectrum is reliable (between 
2.5 - 7 mHz for sunspot \#1, and between 3 - 8.5 mHz for sunspot \#2). 

\noindent From Figure \ref{fig:phase_spectra} we can see that in the range 
from 0 to 2 mHz, the phase spectrum is very noisy (in both cases), 
with a mean value around
zero, indicating that the oscillation modes at photospheric heights
have nothing to do with the same modes observed at chromospheric heights,
i.e. there is no wave propagation in this frequency regime. From 2 up to, 
approximately, 4 mHz, the phase spectrum is not so noisy but values still
remain very near zero, indicating that there is no propagation, but what
we see are standing waves (i.e., waves that are reflected somewhere 
below the level of formation of the Si line).
From 4 mHz on, the phase spectrum shows a clear increasing tendency
meaning that these frequency modes do propagate from the photosphere
reaching the chromosphere at some point.

\subsubsection{The theory}

Just for a reason of completeness and self-containment,
in this subsection we make a brief overview of some basic
models for wave propagation in plane-parallel atmospheres, following
works available in the literature (Ferraro and Plumpton 1958, Souffrin 1972,
Mihalas and Mihalas 1984, B\"unte and Bogdan 1992). 
Beginning with an adiabatic
stratified atmosphere, we will compare it to a non-stratified one, and
after that, to an atmosphere that allows for radiative losses through
Newton's cooling law. We will see that this last case agrees reasonably
well with the observations.

Consider a standard plane-parallel isothermal stratified atmosphere permeated 
by a uniform vertical magnetic field (as in Ferraro and Plumpton, 1958).
If we introduce a small adiabatic perturbation with a frequency $\omega$, 
and study its propagation, in the linear regime
these authors find two independent solutions: an Alfv\`en wave (transversal in nature,
propagating along the field lines) and a sound wave (also propagating along
the field lines, but longitudinal and totally unaware of the presence of
the magnetic field). As we are studying the propagation of longitudinal 
velocity oscillations
along the magnetic field lines, we are only interested in the sound wave.
The amplitude $A(z)$ of the generated monochromatic wave 
appears as the solution to the differential equation

\begin{equation}
c^2 \frac{d^2 A(z)}{dz^2} - \gamma g \frac{dA(z)}{dz} + \omega^2 A(z) = 0,
\label{eq:difeq-adiabatic-sound}
\end{equation}

\noindent where $z$ is the cartesian vertical coordinate, $H_0$ is the 
pressure scale height, $g$ is the gravity 
(assumed constant), $c^2=\gamma g H_0$ the speed of sound and 
$\gamma = c_p / c_v$ the ratio of specific heats, which in the case of an 
adiabatic propagation is strictly equal to $5/3$ for a monoatomic plasma. 
If we introduce the solution $A(z) = e^{i k_z z}$ (where $k_z$ represents the 
vertical wavenumber) in Eq. (\ref{eq:difeq-adiabatic-sound}), we end up with a 
dispersion relation of the form:

\begin{equation}
k_z = \frac{1}{c} ( -i \omega_{ac} \pm \sqrt{\omega^2 - \omega_{ac}^2}),
\label{eq:kw-adiabatic}
\end{equation}

\noindent where

\begin{equation}
\omega_{ac} = \gamma g / 2c
\end{equation}

\noindent is the cut-off frequency. When $k_z$ takes an imaginary value 
($\omega < \omega{ac}$), the solution $A(z)$ is damped and there 
is no wave propagation. On the contrary, when $k_z$ has a real part 
($\omega > \omega_{ac}$), the solution is a purely upward (downward) 
propagating wave that increases
(decreases) its amplitude as it reaches higher (lower) levels of the
atmosphere. This behavior can be inferred from the equations

\begin{eqnarray}
(\omega < \omega_{ac}) \qquad A(z) & = & e^{\frac{\omega_{ac} \pm \sqrt{\omega_{ac}^2 - \omega^2}}{c}z} \\
(\omega > \omega_{ac}) \qquad A(z) & = & e^{z \omega_{ac}/c} e^{\frac{\pm i\sqrt{\omega^2-\omega_{ac}^2}}{c}z}
\label{eq:solutions-adiabatic}
\end{eqnarray}

\noindent Below the cut-off frequency, oscillations do not propagate, 
being instead evanescent in character, and generating standing waves.
In the case of standing waves, the difference in phase of the oscillations
measured at whatever heights are chosen, is always zero.
Above the cut-off value, oscillation modes propagate with a phase
velocity that depends on the frequency. The phase of the oscillation
is the argument of the complex exponential 
($\phi = c^{-1} \sqrt{\omega^2 - \omega_{ac}^2} z$), and the phase 
difference of the oscillations of a propagating wave measured at two heights
will be $\Delta \phi = c^{-1} \sqrt{\omega^2 - \omega_{ac}^2}\Delta z$,
where $\Delta z$ is the geometric distance between the two levels.

\begin{figure}[ht]
\begin{center}
\includegraphics[angle=0,scale=0.5]{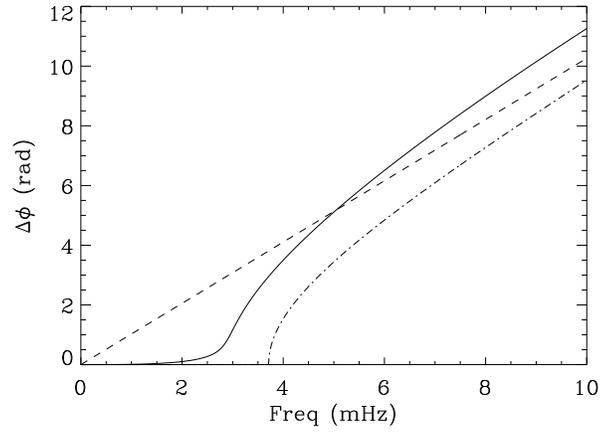}
\caption{{\bf Dashed:} Phase spectrum for linear wave propagation in an isothermal non stratified atmosphere. {\bf Dot-dashed:} the same in a stratified atmosphere (including gravity). {\bf Solid:} The same as the dot-dashed case, but allowing for radiative losses with a Newton's cooling law ($\tau_R = 15$ s). The same values are applied to the common parameters in the three cases (T = 9000 K, $\Delta z$ = 1600 km, g = 274 m s$^{-2}$).}
\label{fig:theor-stratified-losses}
\end{center}
\end{figure}

\noindent The dot-dashed line in Figure \ref{fig:theor-stratified-losses} 
represents the
phase difference of the oscillations measured at two fixed heights for the 
case of linear adiabatic vertical propagation in an isothermal stratified 
atmosphere. 
Below the cut-off frequency ($\sim 3.7$ mHz in this simulation) 
nothing propagates, while  above it, modes start to propagate with a phase 
speed that decreases with frequency. The medium is then dispersive. 
The dashed line (just for comparison) shows the case for linear adiabatic wave 
propagation in an isothermal non-stratified atmosphere (without gravity).
In this case, the phase difference is linear with frequency, meaning that the 
phase velocity is the same for all the oscillation modes,
and that there exists no cut-off frequency - i.e. all modes propagate. This
is the case of a non-dispersive medium.

If, instead of an adiabatic propagation, we relax this condition
allowing for radiative losses with a 
simple Newton's cooling law (following Mihalas and Mihalas 1984, but 
originally developed by Souffrin, 1972), the
picture we obtain is somewhat different. 
Newton's cooling law accounts for the damping of the temperature fluctuations
due to radiative losses, with a typical cooling time $\tau_R$ given by:

\begin{equation}
\tau_R = \rho c_v / (16 \chi \sigma_R T^3),
\end{equation}

\noindent where $\chi$ is the mean absorption coefficient and $\sigma_R$ is
the Stefan-Boltzmann constant. 
We can use their solution for vertical propagation of longitudinal 
waves (i.e. zero horizontal wavenumber $k_x = 0$) in the case of the
propagation of acoustic-gravity waves in a radiating fluid for a
non-magnetic isothermal atmosphere. 
The reason that allows us to do this, is based on the fact that sound waves 
propagating along vertical magnetic field lines are unaware of the presence 
of the magnetic field. This means that the differential equation for sound 
waves 
propagating along a vertical magnetic field will be formally identical to the 
one for the field free case. The solution inserted into the differential
equation

\begin{equation}
A(z) = D e^{z/(2H_0)} e^{i k_z z}
\end{equation}

\noindent yields to the following dispersion relation:

\begin{equation}
k_z^2 = \frac{\omega^2-\hat\omega_{ac}^2}{\hat c^2},
\end{equation} 

\noindent where, following the definitions by B\"unte and Bogdan (1992)

\begin{equation}
\hat \omega_{ac} = \hat c^2 / 4 H_0; \qquad \hat c^2 = \hat \gamma g H_0; \qquad \hat \gamma = \frac{1 - \gamma i \omega \tau_R}{1 - i\omega \tau_R}.
\end{equation}

\noindent We can compute the real and the imaginary parts of $k_z$:

\begin{eqnarray}
k_R^2 & = & \frac{1}{2}[h_R + (h_R^2+h_I^2)^{1/2}],\\
k_I^2 & = & \frac{1}{2}[-h_R +(h_R^2 + h_I^2)^{1/2}],
\end{eqnarray}

\noindent where

\begin{eqnarray}
h_R & = &\frac{\omega^2(1+\omega^2\tau_R^2\gamma)}{gH_0(1+\omega^2\tau_R^2\gamma^2)}-\frac{1}{4H_0^2}\\
h_I & = &\frac{(\gamma - 1)\tau_R\omega^3}{(1+\omega^2\tau_R^2\gamma^2)g H_0}.
\end{eqnarray}

\noindent Both $k_R$ and $k_I$ are real, and the curves $h_R = 0$ define
the boundaries between mainly propagating ($k_R > k_I$) and mainly damped 
($k_R < k_I$) waves. The solid line in Figure 
\ref{fig:theor-stratified-losses} shows us what the phase spectrum would
look like in this case. Now, there is no cut-off 
frequency as such, being all modes propagated and 
reflected at the same time, with a ratio of propagation versus reflection 
increasing as a function of frequency. The transition between propagating 
and non-propagating regimes is not so clear though a pseudo-cut-off 
frequency can be defined. When the
typical radiative time scale $\tau_R$ is small enough (of the order of tens 
of seconds),
this effective cut-off frequency turns out to be much smaller than the 
one obtained for the adiabatic case.

\subsubsection{Combining theory and observations}

We were able to fit the phase spectra with this last asumption (stratified
atmosphere allowing for radiative losses) which leaves three free parameters:
the temperature of the model atmosphere $T$, the difference in heights 
$\Delta z$ at which the oscillations are measured, and the typical time scale 
in which the temperature fluctuations are damped radiatively, $\tau_R$.

\begin{figure}
\begin{center}
\includegraphics[angle=0,scale=0.4]{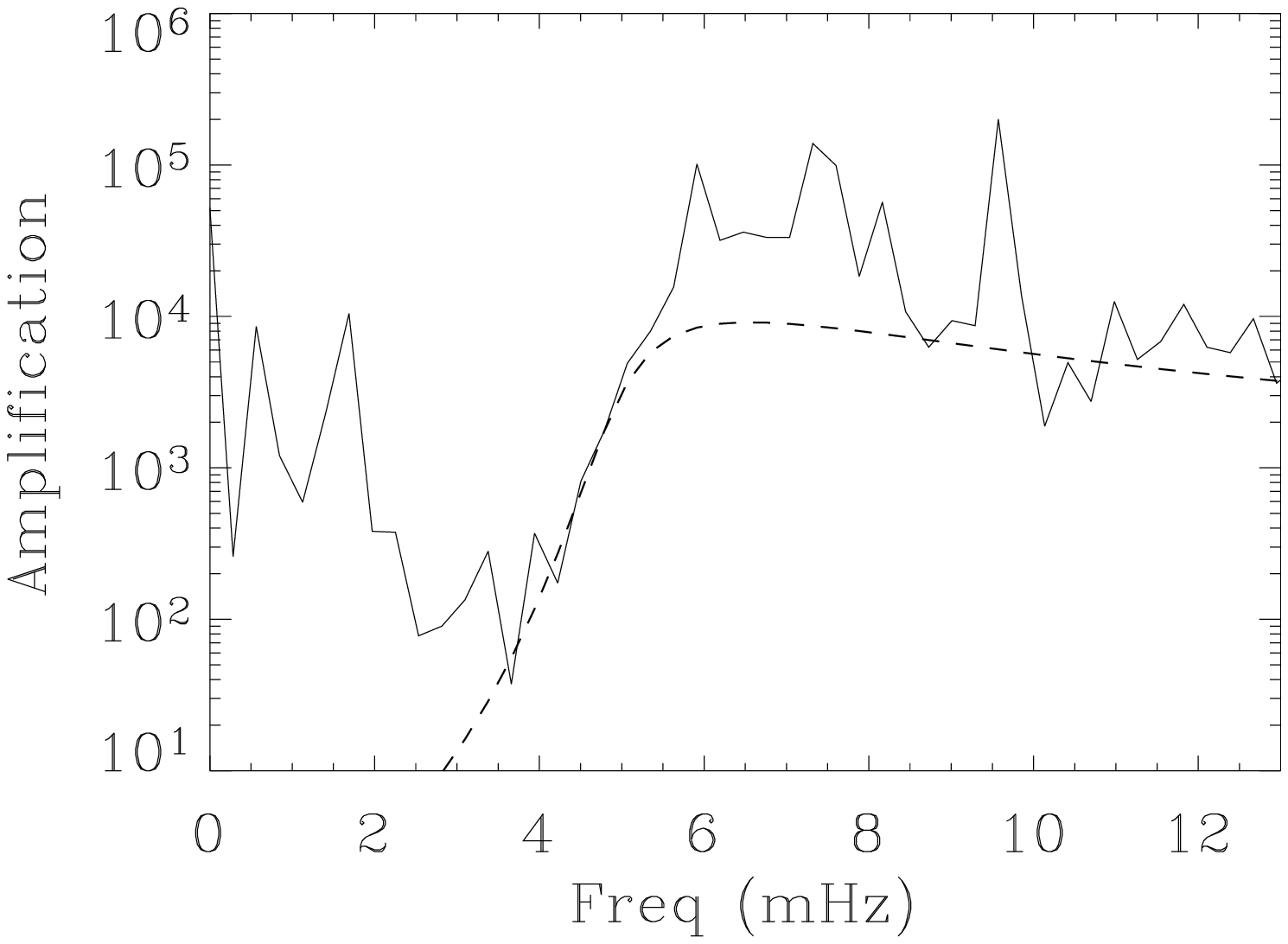}\\
\includegraphics[angle=0,scale=0.4]{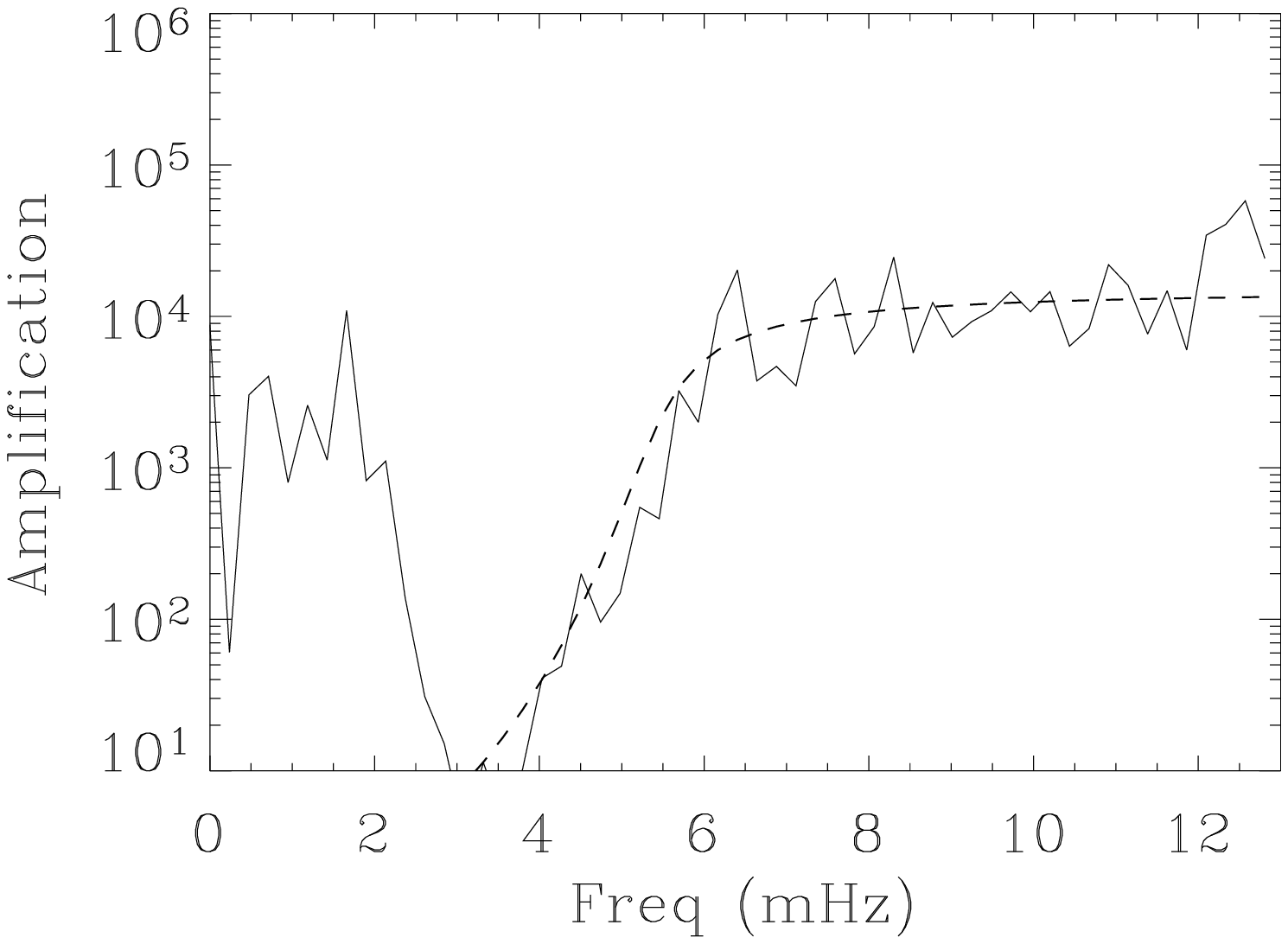}
\caption{{\bf Solid:} Ratio of chromospheric over photospheric power as a function of frequency for data sets 1 (top) and 2 (bottom). {\bf Dashed:} best fit from the model.} 
\label{fig:amplification}
\end{center}
\end{figure}

\noindent The model accounts for the
effective cut-off frequency, the slow transition between the propagating
and non-propagating regimes, and the slope of the phase spectra above
the atmospheric cut-off. The solid line in Figure \ref{fig:phase_spectra} shows
the best fit for both data sets. The values used for the fits are
detailed in Table \ref{table:fits} and will be discussed in section 5. 
The model accounts not only for the phase
spectra, but also for the amplification factor of the chromospheric signals 
relative to their photospheric counterpart, as a function of frequency. 
Figure \ref{fig:amplification} shows the ratio of chromospheric
over photospheric power as a function of frequency for both data sets.
Overplotted to the observational ratio, we show the theoretical one 
(in dashed line)
obtained from the best-fit-parameters applied to the model. Below 3 mHz
the data are not reliable due to the S/N ratio, but note that
above this value, the agreement is quite good (in tendency and order of
magnitude).
\noindent Authors before have tried to fit phase spectra with the 
non-stratified model, which cannot even reproduce a cut-off frequency, not
to talk about the change in slope of the phase spectrum.  
Taking into account that the atmospheric model we use is very simple 
(isothermal and linear), the fits agree reasonably well with the observations 
and account for the main features.

\subsection{Filtering}

In order to determine how the photospheric power spectrum 3-minute peak 
is related to the chromospheric oscillation, we filter the velocity signals in 
narrow bands around 6 mHz (where power is significant at
both heights) for
each point inside the slit. This allows us to compare the 
photospheric and chromospheric filtered velocity maps and see what
time shift we have to apply between them so that they 
match each other. 

\noindent From the curve that fits the phase spectrum, we can easily obtain the
group velocity ($v_g = d\omega / dk$), and from that, the time that it would 
take for each oscillating mode to reach the chromosphere from the photosphere
($t_{delay} = \Delta z / v_{g}$).

\begin{figure}
\begin{center}
\includegraphics[angle=0,scale=0.4]{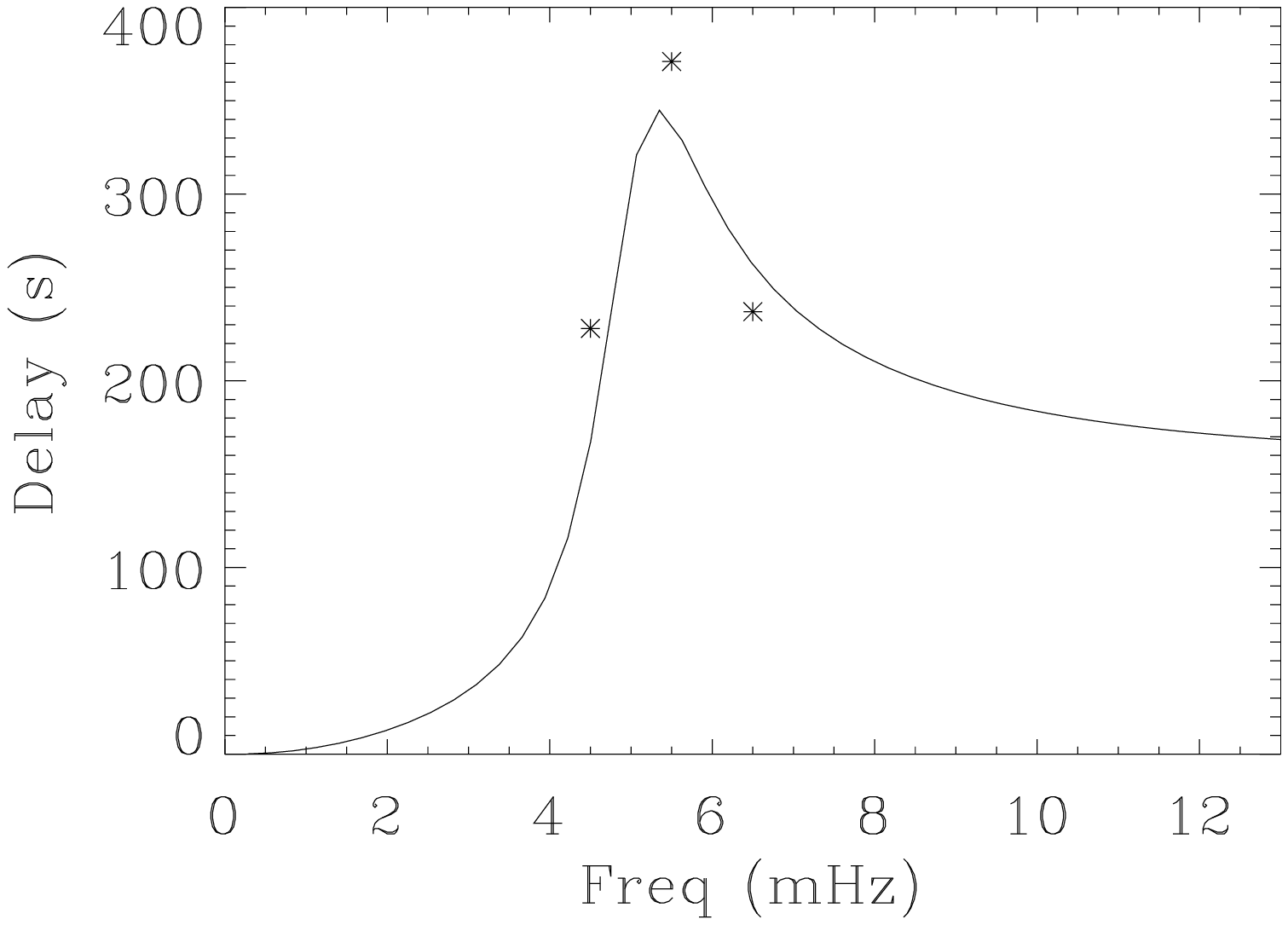}\\
\includegraphics[angle=0,scale=0.4]{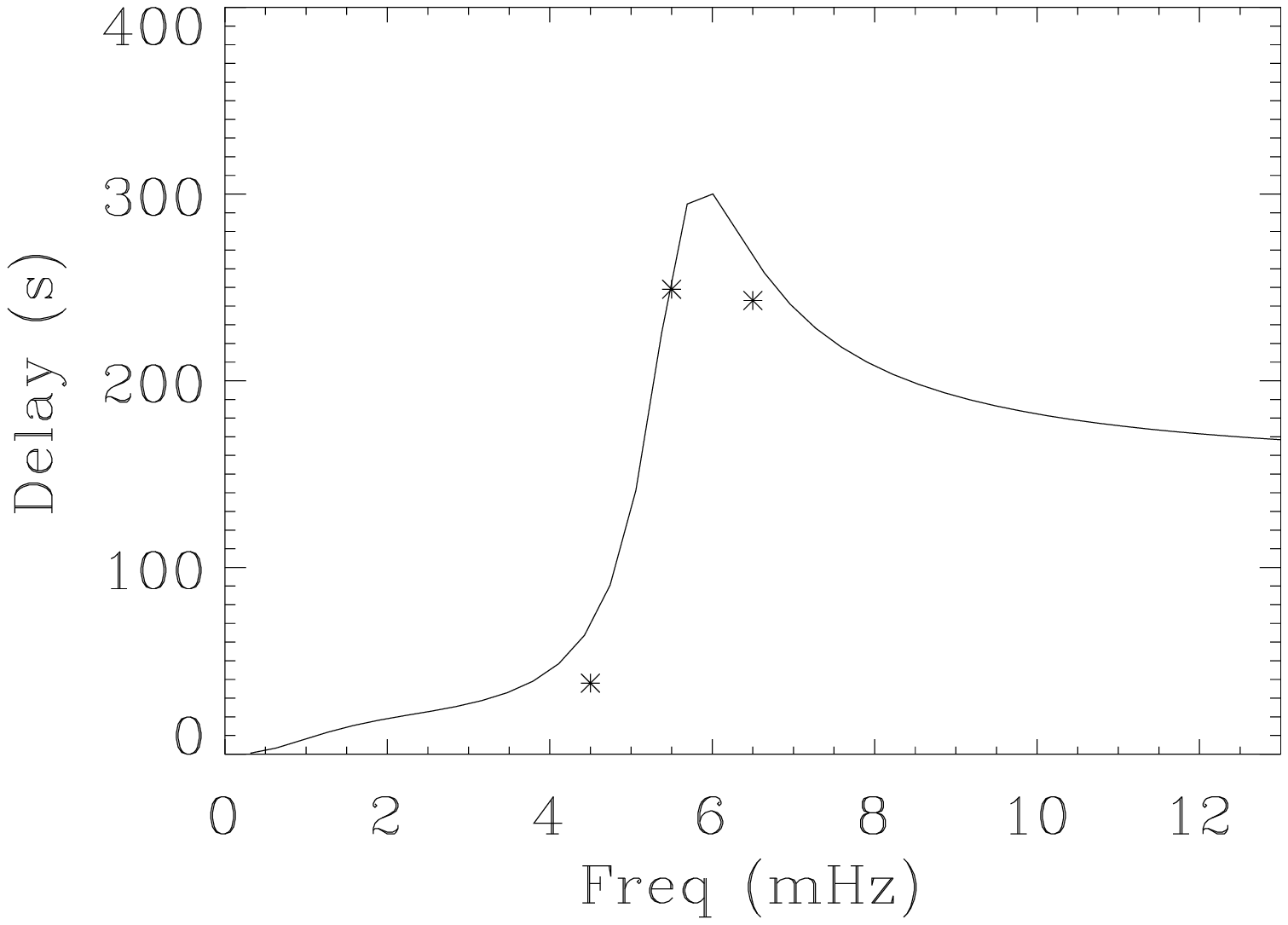}
\caption{Solid line represents the time that it would take for a quasi-monochromatic photospheric perturbation to reach the chromosphere, as a function of frequency, obtained directly from the fits to the phase spectra (sunspot \#1 on the top panel and sunspot \#2 on the bottom one). Stars represent the measured values of the time delay within 3 narrow filtering bands.}
\label{fig:obs-theor-time-delay}
\end{center}
\end{figure}

Figure \ref{fig:obs-theor-time-delay} shows the time (solid line) 
that a quasi-monochromatic photospheric 
perturbation would take to reach the chromosphere, calculated directly from
the fit to the phase spectra of both data sets. The phase velocity, and 
consequently the time delay, is highly dependent on frequency. 
This implies that, in order to estimate the time that a perturbation 
originated in the low photosphere takes to reach the high chromosphere, we 
should compare the modulation pattern of the velocity signals filtered 
in narrow frequency bands, so that the propagating time does not vary 
significantly within the filtering range.

\noindent We take both the photospheric and the chromospheric line of sight 
velocity maps and we filter them in three narrow frequency bands: 4 - 5, 5 - 6 
and 6 - 7 mHz. We are not interested in the signals below 4 mHz since from the 
phase spectrum we see that there is no significant wave propagation in this 
range. Above 7 mHz, the phase spectrum becomes very noisy, and the signal
in the power spectrum too low to be trusted.
After comparing each pair of maps filtered in the same frequency 
range, we find that the external photospheric and chromospheric modulation 
patterns resemble each other, but, in order to make them match, a temporal 
shift has to be applied between them.

\begin{figure}
\begin{center}
\includegraphics[angle=0,scale=0.5]{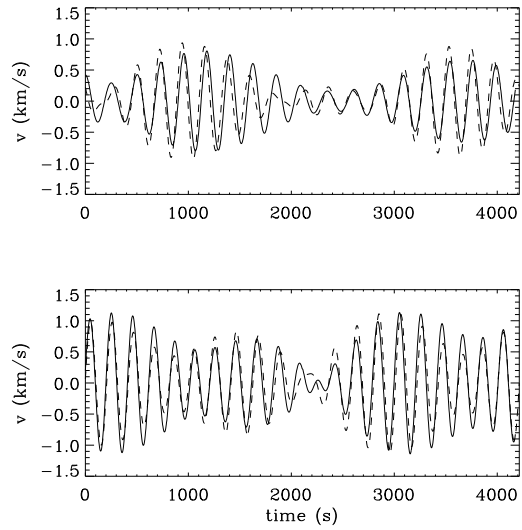}
\caption{{\bf Solid:} Chromospheric velocity signal filtered in the 4 to 5 mHz band for two positions (upper and lower panels) inside umbra \#2. {\bf Dashed:} Photospheric filtered velocity signal (in the same band) for the corresponding positions inside the umbra. Photospheric velocity has been amplified by a factor of 20 and shifted forward 38 seconds in order to match the chromospheric one.}
\label{fig:filtered1}
\end{center}
\end{figure}

\begin{figure}
\begin{center}
\includegraphics[angle=0,scale=0.5]{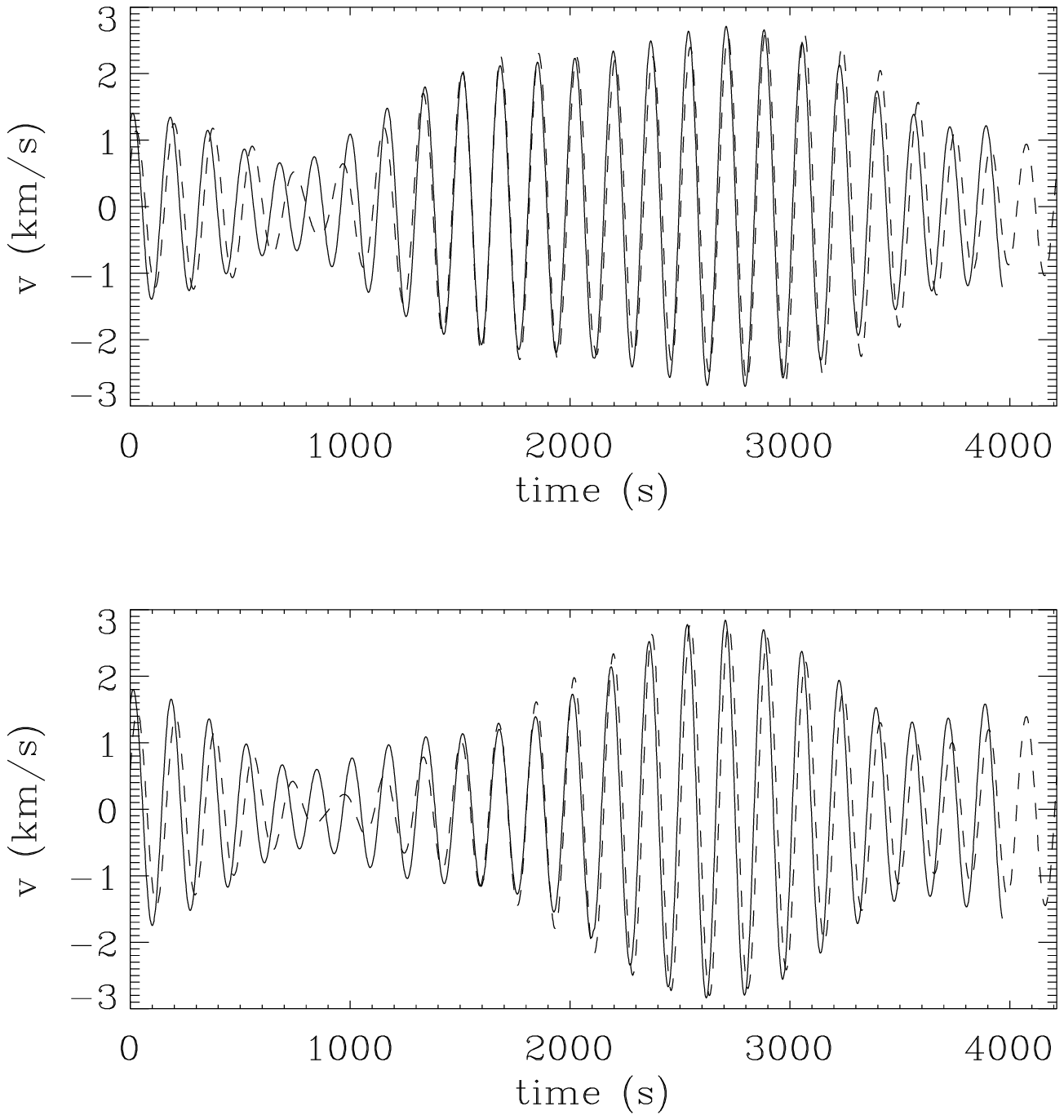}
\caption{Analogous to Fig \ref{fig:filtered1} but in the 5 to 6 mHz range. Photospheric velocity has been amplified by a factor of 80 and shifted forward 242 seconds in order to match the chromospheric one.}
\label{fig:filtered2}
\end{center}
\end{figure}

\begin{figure}
\begin{center}
\includegraphics[angle=0,scale=0.5]{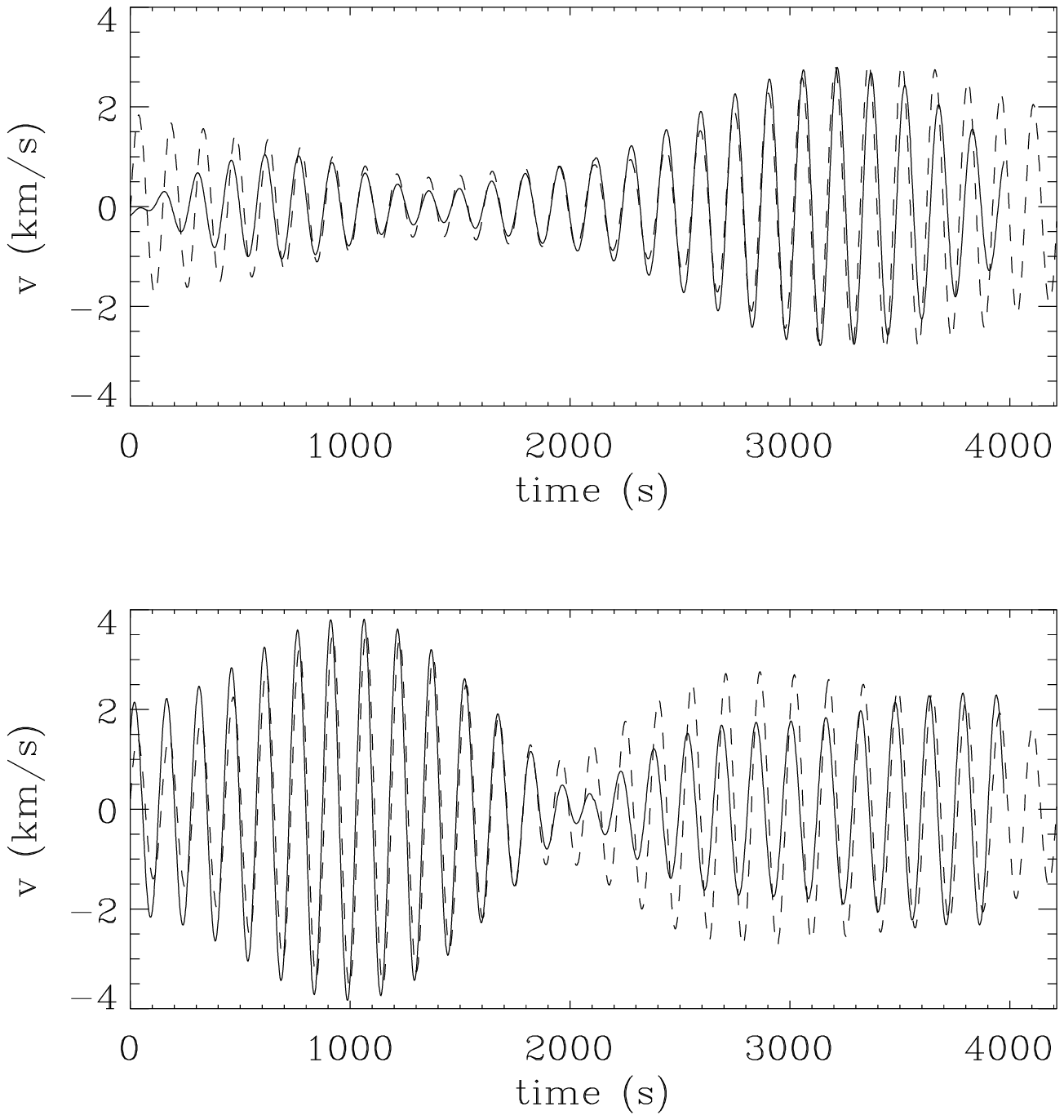}
\caption{Analogous to Fig \ref{fig:filtered1} but in the 6 to 7 mHz range. Photospheric velocity has been amplified by a factor of 80 and shifted forward 248 seconds in order to match the chromospheric one.}
\label{fig:filtered3}
\end{center}
\end{figure}

Figure \ref{fig:filtered1} shows the filtered chromospheric (solid) and
photospheric (dashed) velocity signals in the 4 - 5 mHz range, for two
positions (upper and lower panels) inside umbra \#2. 
We have shifted the photospheric velocity signal with respect to the 
chromospheric one in order to achieve a correspondence between the modulation 
patterns, yielding a time delay of roughly 40 seconds. The sense of 
the shift is such that what happens in the photosphere comes before the 
corresponding chromospheric events - i.e. upward propagation.
Figures \ref{fig:filtered2} and \ref{fig:filtered3} are completely analogous
to Figure \ref{fig:filtered1}, but filtered in the 5 - 6 and 6 - 7 mHz bands 
respectively. The time delay we had to apply was 242 s in the first
case, and 248 s in the second one.

\noindent Stars overplotted to the theoretical time delay in Fig.
\ref{fig:obs-theor-time-delay} correspond to the temporal shift we had
to apply between photospheric and chromospheric filtered velocity maps in 
order to make them match. Even though the theoretical curve predicts
a strong variation of the time delay within the 1 mHz filtering bands, 
the agreement between theory and observations is pretty good.

The measured time delay remains constant
along the slit within each filtering frequency band. We find that the match in 
the shape of the modulation schemes of the 
photospheric and chromospheric signals is not just a coincidence, but
remains along most of the umbra (for both sunspots analyzed in this paper).
Only the edges of the umbra show missmatch between the signals. 
This is a consequence of the magnetic field 
lines opening up as they approach higher layers in the solar atmosphere.
Wave propagation is taking place along these field lines, implying
that the modulation pattern of the oscillation broadens as the
waves propagate higher in the atmosphere, so we will not see the edges of
the photospheric map in the chromospheric one.

\section{Discussion}

Many authors before us have given evidence for the upward propagation of 
waves in the solar atmosphere.
For instance, Brynildsen et al. (2003, 2004) talk about propagation from 
the upper chromosphere to the transition region and into the corona, 
based on spectroscopic observations made with TRACE and SUMER of transition 
region lines. They find that the 3 min oscillations are easier to measure in 
the blue wings of these lines than in the red wings, giving support to
the hypothesis of upwardly propagating acoustic disturbances. 

In this paper, we study the relationship of the line of sight (LOS) velocity 
signals at the low photosphere and the high chromosphere inside umbral
atmospheres by means of spectropolarimetry in the near-IR spectral
region around 10830 \AA. 
Photospheric power spectra show a non-negligible amount of power
in the 6 mHz band (3 min oscillations) within the umbrae of both sunspots.
Analyzing the LOS velocity phase spectra for both data sets, we find that
the power above 4 mHz indeed reaches the chromosphere, while for lower 
frequencies the energy does not seem to propagate up to those heights.
We managed to fit the phase and amplification spectra with a simple model 
of linearized vertical wave propagation in a stratified atmosphere which 
allows for radiative losses. The model accounts for the
effective cut-off frequency, the slow transition between the propagating
and non-propagating regimes, the slope of the phase spectra above
the $\omega_{ac}$ and the amplification factor of chromospheric versus
photospheric power as a function of frequency.
Taking into account the simplicity of the model (which depends only 
on three free parameters), we cannot expect very realistic numbers for
the physical magnitudes yielded from the fits. But, interestingly, the
retrieved values seem to be somewhat coherent with what one would 
expect. The height difference between the layers of formation of the 
Silicon and de Helium lines is the same for both sunspots. On the 
other hand, the temperature is lower inside the biggest sunspot, in 
agreement with Collados et al. (1994) and Maltby (1992).
Another issue is the value obtained for the typical radiative relaxation time, 
which is smaller for the largest sunspot (\#1). If we assume a monolitic
model for sunspot umbrae, we would expect 
larger values of $\tau_R$ the more homogeneus the structure. 
Smaller values of $\tau_R$ should be related to larger temperature 
inhomogeneities, what, at first sight, is incompatible with our results. 
However, this disagreement could be explained if we take into account that 
sunspot \#1, although larger, was actually divided by a faint light bridge, 
being more inhomogeneus in nature than sunspot \#2.

\noindent The model also gives us the time delay we should expect between 
photospheric and chromospheric signals, which happens to be extremely 
dependent on the frequency of the propagating mode. 
In order to compare this with the data, we filtered the velocity maps in 
three narrow frequency bands: 4 - 5, 5 - 6 
and 6 - 7 mHz. Then we compared each pair of photospheric and chromospheric 
maps filtered in the same frequency range, finding that the external modulation
patterns resemble each other. This shows clearly the temporal shift we have 
to apply to one of the maps in order to make it match the other one. 
This method yields to values for the time delay that agree reasonably well 
with the prediction by the theoretical model.

All along this paper we have assumed linear propagation of waves disregarding
the effects of non-linear terms in the MHD equations, although in our data sets
there is clear evidence for non-linearity of the velocity oscillations at 
chromospheric heights (i.e. shock waves). Is the linear approximation still 
valid in this case?

\noindent Many authors before (see e.g. Fleck and Schmitz 1993) 
have argued about the possibility of 
non-linear interaction among the 5-minute (or even high frequency) 
modes being the source of the 
three minute oscillations. The non-linear terms
in the MHD equations would take some power from the 5-minute band and
convert it into higher frequency modes, thus giving rise to the 3-minute
oscillations.
If, in between the Silicon and the Helium levels of formation this were 
the case, the five-minute band would be contributing to the generation of the 
three minute power along all the distance in between photosphere and the 
high chromosphere, and so the chromospheric signal would not resemble the 
photospheric one at all at the same frequency. This does not 
agree with the fact that there exists a clear correlation between 
photospheric and chromospheric velocity maps in the $6$ mHz frequency range.
We can conclude that, in between the levels of formation of the Silicon and
the Helium lines, the non-linear interaction among the 5-minute modes is
not the main cause of the 3-minute power. Of course there can still be 
some of this taking place, but the main contribution to the chromospheric
oscillations comes directly from the photosphere. 

\noindent On the other hand, the saw-tooth shape of the chromospheric velocity 
signals is related to high frequency terms in the chromospheric power 
spectrum. These high frequency modes appear probably as a consequence of 
non-linear interaction among 3-minute modes, since the amplitude of 
chromospheric 
oscillations is such that the linear assumption cannot be made anymore. 
Our study is restricted to the region of the spectra below $8$ mHz,
so we are ignoring these high frequency modes, and with them, the 
effects of non-linearity. 

\noindent A different but related issue is the final origin of the 3-minute 
power at photospheric heights, which may come from non-linear processes or 
have a completely independent cause. The information we get from our data
cannot address this still open question.
It seems clear to us that the oscillation pattern measured in the 
chromosphere can clearly be seen at photospheric layers. 
Wherever (or whatever) the source is, at some point, the 3-minute mode gets
to the photosphere and several minutes later reaches the high chromosphere.

\section{Conclusions}

The spectropolarimetric investigation we have presented here
provides observational evidence for the upward propagation of
slow magneto-acoustic waves 
from the photosphere to the high chromosphere inside the umbra of a 
sunspot. 
The time delay between the signals corresponding to both regions varies
strongly with the frequency of the oscillation, going from a few tens of
seconds to several minutes.
As the photospheric perturbations propagate upwards, their amplitude increases 
due to the rapid decrease in density, and they eventually develop into shock 
waves at chromospheric heights.

Interestingly, 
the observed temporal variability of the Stokes profiles 
in the Si {\sc i} line at 10827.09 \AA\ may help to establish the
required initial condition for performing
realistic MHD simulations, which
are needed for a full physical understanding of the
phenomenon of wave propagation in sunspot atmospheres.
Our future work on this topic will focus on similar spectropolarimetric
investigations, but for atmospheric plasma structures with
lower manetic fluxes, such as pores, active region plages and the
chromospheric network of the `quiet' Sun.

It is a pleasure to thank H\'ector Socas-Navarro and Jos\'e Antonio Bonet 
for fruitful discussions and for sharing with us their knowledge on Stokes 
inversion techniques and Fourier analysis. We are also grateful to Bruce 
Lites and Tom Bogdan for suggesting improvements to an earlier version of this 
paper.
This research has been funded by the Spanish Ministerio de 
Educaci\'on y Ciencia through the proyect AYA2004-05792,
and is part of the European Solar Magnetism Network.

\clearpage

\clearpage

\clearpage

\clearpage

\clearpage

\clearpage

\clearpage

\clearpage

\clearpage

\clearpage

\clearpage

\clearpage

\clearpage

\begin{table}
\caption{Details of data sets.\label{tbl-1}}
\begin{center}
\begin{tabular}{c | ccccc}
\tableline\tableline
Data set & date  & location & umbral size (") & sampling (s) & duration (s)\\
\tableline
1  &  Oct 1st, 2000 & 11S 2W   & 16 &  7.9  & 3555 \\
2  &  May 9th, 2001 & 20N 25W  & 10 &  2.1  & 4200 \\
\tableline
\end{tabular}
\end{center}
\end{table}

\clearpage

\begin{table}
\caption{Parameters for phase spectra fits.\label{table:fits}}
\begin{center}
\begin{tabular}{c | ccc}
\tableline\tableline
Data set & T (K) & $\Delta z$ (km) & $\tau_R$ (s)\\
\tableline
1  &  3400 & 1000 & 13 \\
2  &  4000 & 1000 & 55 \\
\tableline
\end{tabular}
\end{center}
\end{table}

\end{document}